\documentclass{article}

\usepackage[utf8]{inputenc} % allow utf-8 input
\usepackage[T1]{fontenc}    % use 8-bit T1 fonts
\usepackage{hyperref}       % hyperlinks
\usepackage{url}            % simple URL typesetting
\usepackage{booktabs}       % professional-quality tables
\usepackage{amsfonts}       % blackboard math symbols
\usepackage{nicefrac}       % compact symbols for 1/2, etc.
\usepackage{microtype}      % microtypography

\usepackage{amsmath,amsfonts,amsthm,amssymb}

\newtheorem{propo}{Proposition}[section]
\newtheorem{lemma}[propo]{Lemma}

\newtheorem{thm}{Theorem}

\newtheorem{ansatz}{Ansatz}

\usepackage{graphicx,float,wrapfig}
\usepackage{multirow}

\def\hf{\widehat{f}}
\def\cL{{\cal L}}
\def\ha{\widehat{a}}

\def\reals{{\mathbb R}}
\def\Z{{\mathbb Z}}
\def\<{\langle}
\def\>{\rangle}
\def\E{{\mathbb E}}
\def\ind{{\mathbb I}}
\def\hH{{\widehat{H}}}
\def\hI{{\widehat{I}}}
\def\cS{{\cal T}}
\def\bh{h}
\def\tS{{\tilde{S}}}
\def\tA{{\widetilde{A}}}
\def\Pr{{\mathbb P}}
\def\tSigma{{\widetilde{\Sigma}}}
\def\tB{{\widetilde{B}}}
\def\ta{{\widetilde{a}}}
\def\Var{{\textrm{ Var }}}

\title{
Breaking the Bandwidth Barrier: \\Geometrical Adaptive Entropy Estimation
}

\author{
 Weihao Gao\thanks{Coordinated Science Lab and Department of Electrical and Computer Engineering}, \;\;
 Sewoong Oh\thanks{Coordinated Science Lab and Department of Industrial and Enterprise Systems Engineering}, \;\;
 and \;\;Pramod Viswanath$^*$\\
   University of Illinois at Urbana-Champaign\\
  Urbana, IL 61801  \\
  \texttt{\{wgao9,swoh,pramodv\}@illinois.edu} \\
}
\date{}

\begin{document}
% \nipsfinalcopy is no longer used

\maketitle

\begin{abstract}

Estimators of information theoretic measures such as entropy and mutual information  are a basic workhorse for many downstream applications in modern data science. State of the art approaches have been either geometric (nearest neighbor (NN) based) or kernel based
(with a globally chosen bandwidth).
%to be data independent and vanishing sub linearly in the sample size).
%(with bandwidth chosen to be data independent and vanishing sub linearly in the sample size).
In this paper, we combine both these approaches to design new estimators of entropy and mutual information that outperform state of the art methods.
%In this paper we  combine both these approaches to design new estimators of entropy and mutual information by borrowing the good ideas of both approaches and strongly outperforms
% all state of the art methods.
 Our estimator uses local bandwidth choices of %{\em
 $k$-NN distances with a finite $k$, independent of the sample size.
Such a local and data dependent choice improves performance in practice, but
the bandwidth is  vanishing at a fast rate, leading to a non-vanishing bias.
%necessitating a  bias cancellation.
We show that the asymptotic bias of the proposed estimator is
{\em universal}; it is independent of the underlying distribution.
Hence, it can be precomputed and subtracted from the estimate.
As a byproduct, we obtain a unified way of obtaining {\em both} kernel and NN estimators.
 The corresponding theoretical contribution relating the asymptotic geometry of nearest neighbors  %NN statistics distances
 to order statistics  is of independent mathematical interest.

\end{abstract}

\section{Introduction}

%\section{Introduction}
Unsupervised representation learning is one of the major themes of modern data science; % and a long studied topic under machine learning.  While there are several approaches, %\cite{}, need citations here
a common theme among the various approaches is to extract  maximally ``informative" features  via {\em information-theoretic metrics} (entropy, mutual information and their variations) -- the primary reason for the popularity of information theoretic measures is that they are invariant to one-to-one transformations and that they obey natural axioms such as data processing. Such an approach is evident in many applications, as varied as computational biology \cite{krishnaswamy2014conditional}, sociology \cite{reshef2011detecting} and information retrieval \cite{manning2008introduction}, with the citations  representing a mere  smattering of recent works. Within mainstream machine learning,
a systematic effort at unsupervised clustering and hierarchical information extraction is conducted in recent works of \cite{ver2014discovering,steeg2015information}. The basic workhorse in all these methods is the  computation of mutual information (pairwise and multivariate) from i.i.d.\ samples. Indeed, {\em  sample-efficient estimation} of mutual information emerges as the central scientific question of interest in a variety of applications, and is also  of fundamental interest to statistics, machine learning and information theory communities.

While these estimation questions have been studied in the past three decades (and summarized in \cite{wang2009foundations}), the renewed importance  of estimating information theoretic measures in a {\em sample-efficient} manner is persuasively argued in a recent work \cite{GVG14}, where the authors note that existing estimators perform poorly in several key scenarios of central interest (especially when the high dimensional random variables are strongly related to each other). The most common estimators (featured in scientific software packages) are nonparametric and involve $k$ nearest neighbor (NN) distances between the samples.  The widely used estimator of mutual information is the one by Kraskov and St\"{o}gbauer and Grassberger  \cite{KSG04} and christened  the KSG estimator   (nomenclature based on the authors, cf.\ \cite{GVG14}) -- while this estimator works well in practice (and performs much better than other approaches such as those based on kernel density estimation procedures), it still suffers in high dimensions. The basic issue is that the KSG estimator (and the underlying differential entropy estimator based on nearest neighbor distances by Kozachenko and Leonenko (KL) \cite{KL87}) does not take advantage of the fact that the samples could lie  in a smaller dimensional  subspace (more generally, manifold) despite the high dimensionality of the data itself. Such lower dimensional structures effectively act as boundaries, causing the estimator to suffer from what is known as boundary biases.

Ameliorating this deficiency   is the central theme of recent works \cite{GVG15,GVG14,lombardi2016nonparametric}, each of which aims to improve upon the classical  KL (differential) entropy estimator of \cite{KL87}.  A local SVD is used to heuristically improve the density estimate
at each sample point in \cite{GVG14}, while a local Gaussian density (with empirical mean and covariance weighted by NN distances)   is heuristically used for the same purpose in \cite{lombardi2016nonparametric}. Both these approaches, while inspired and intuitive,  %constitute a systematic effort  at addressing the key issues at play; in particular, no
 come with no theoretical guarantees (even consistency) and from a practical perspective involve delicate choice of key hyper parameters. An effort towards a systematic study is initiated in \cite{GVG15} which connects the aforementioned heuristic efforts of \cite{GVG14,lombardi2016nonparametric} to the {\em local log-likelihood} density estimation methods \cite{HJ96,Loa96} from theoretical statistics.

The local density estimation method is a strong generalization of the traditional kernel density estimation methods, but requires a delicate normalization which necessitates the solution of certain integral equations (cf.\ Equation (9) of \cite{Loa96}). Indeed, such an elaborate numerical effort is one of the key impediments for the entropy estimator of \cite{GVG15} to be practically valuable. A second key impediment is that theoretical guarantees (such as consistency) can only be provided when the bandwidth is chosen globally %independent of the samples
 (leading to poor sample complexity in practice) and consistency requires the bandwidth $h$ to be chosen such that
$n h^d \to \infty$ and $h\to 0 $, where $n$ is the sample size and $d$ is the dimension of the random variable of interest.
%{vanishingly small as the sample size increases (leading to choice of hyper parameters governing the sublinear rate of decay).}
More generally, it appears that a systematic application of local log-likelihood methods to
estimate
%use samples to compute
{\em functionals} of the unknown density from i.i.d. samples is missing in the theoretical statistics literature (despite local log-likelihood methods for regression and density estimation being standard textbook fare \cite{Was06,loader2006local}). We resolve each of these deficiencies in this paper by undertaking a  comprehensive study of estimating the (differential) entropy and mutual information from i.i.d.\ samples using sample dependent bandwidth choices (typically {\em fixed} $k$-NN distances). This effort allows us to connect disparate threads of ideas from seemingly different arenas: NN methods, local log-likelihood methods, asymptotic order statistics and sample-dependent heuristic, but inspired, methods for mutual information estimation suggested in the work of \cite{KSG04}.  %Our {\bf Main Results} are enumerated below. %We enumerate our main contributions below.

{\bf Main Results}: We make the following contributions.
\begin{enumerate}
\item {\bf Density} estimation: Parameterizing the log density by a polynomial of degree $p$, we derive {\em simple closed form} expressions for the local log-likelihood maximization problem for the cases of $p\leq 2$ for arbitrary dimensions, with Gaussian kernel choices.
This derivation,  posed as an exercise in  \cite[Exercise 5.2]{loader2006local}, significantly improves the computational efficiency upon similar endeavors in the recent efforts of \cite{GVG15,lombardi2016nonparametric,vincent2003locally}.

\item {\bf Entropy} estimation: Using resubstitution of the local density estimate, we derive a simple closed form estimator of the  entropy using a sample dependent bandwidth choice (of $k$-NN distance, where $k$ is a {\em fixed} small integer independent of the sample size): this estimator  outperforms  state of the art entropy estimators in a variety of settings.  Since the bandwidth is data dependent and vanishes too fast (because $k$ is fixed), the estimator has a bias, which we derive a closed form expression for and show that it is {\em independent} of the underlying distribution and hence can be easily corrected:
this is our main theoretical contribution, and involves new theorems
on asymptotic statistics of nearest neighbors generalizing classical work in probability theory \cite{Rei12}, which might be of independent mathematical interest.
%In classical work in probability theory \cite{Rei12}, it is known that the enarest neighbor distances converge to

\item {\bf Generalized} view: We show that seemingly very different approaches to entropy estimation -- recent works of \cite{GVG14,GVG15,lombardi2016nonparametric} and the classical work of fixed $k$-NN estimator of Kozachenko and Leonenko \cite{KL87} -- can all be cast in the local log-likelihood framework as specific kernel and {\em sample dependent bandwidth} choices. This allows for a unified view, which we theoretically justify by showing that resubstitution entropy estimation for {\em any} kernel choice using fixed $k$-NN distances as bandwidth involves a bias term that is {\em independent of the underlying distribution} (but depends on the specific choice of kernel and parametric density family). Thus our work is a strict mathematical generalization of the classical work of \cite{KL87}.

\item {\bf Mutual Information} estimation: The inspired work of \cite{KSG04} constructs a mutual information estimator that subtly altered (in a sample dependent way) the three KL entropy estimation terms, leading to superior empirical performance. We show that the underlying idea behind this change can be incorporated in our framework as well, leading to a novel mutual information estimator that combines the two ideas and  outperforms state of the art estimators in a variety of settings.

\end{enumerate}

In the rest of this paper we describe these main results, the sections organized in roughly the same order as the enumerated list.

%Comparisons to KDE and CMU \cite{Joe89,PPS10,PXS12,KKPW15}

%Comparisons to $k$NN estimators

%-- KL estimator \cite{KL87} and KSG \cite{KSG04}

%-- USC: A series of heuristics have been proposed recently in \cite{GVG14,GVG15} for %$k$NN estimators,
%by detecting boundaries and modifying the kernel to mitigate the bias.

%-- cite Verdu \cite{WKV09,WKV09survey}

%-- our work \cite{GKOV16,GOV16}

% ---------------------------------------------------------------------------------------------------------------------------------
\section{Local likelihood density estimation (LLDE)}
\label{sec:llde}
Given  $n$ i.i.d.\ samples $X_1,\ldots ,X_n$, estimating the unknown density $f_X(\cdot)$ in ${\mathbb R}^d$ is a very basic statistical task.  Local likelihood density estimators \cite{Loa96,HJ96} constitute state of the art and
 are specified by  a
weight function $K:\reals^d\to\reals$ (also called a kernel),  a degree  $p\in\Z^+$ of the polynomial approximation,
and the bandwidth $h\in\reals$, and
%to a novel density estimator
maximizes the local log-likelihood:
\begin{eqnarray}
	\label{eq:locallikelihood}
	\cL_x(f) &=&  \sum_{j=1}^n K\left(\frac{X_j-x}{h}\right)\log f(X_j) - n \int K\left(\frac{u-x}{h}\right) f(u) \, du\;,
\end{eqnarray}
%where $K:\reals^d \to \reals$ is an appropriately chosen
%weight function, also called a  kernel, and $h\in\reals$ is a bandwidth.
%The choice of the bandwidth is crucial both in theory and practice, which is  the main focus of this paper.
where maximization is  over an exponential polynomial family, locally approximating $f(u)$ near $x$:
\begin{eqnarray}
	\label{eq:localdensity}
	\log_e  f_{a,x}(u) = a_0 + \<a_1,u-x\> + \<u-x,a_2(u-x)\> + \cdots + a_p[u-x,u-x,\ldots,u-x]\;,
\end{eqnarray}
parameterized by $a=(a_0,\ldots,a_p)\in\reals^{1\times d\times d^2\times\cdots\times d^p}$,
where $\<\cdot,\cdot\>$ denotes the inner-product and
$a_p[u,\ldots,u]$ the $p$-th order tensor projection.
The {\em local likelihood density estimate} (LLDE) is defined as
$\hf_n(x) = f_{\ha(x),x}(x) = e^{\ha_0(x)}$, where
  $\ha(x) \in \arg\max_a  \cL_x(f_{a,x})$.
The maximizer is represented by a series of   nonlinear  equations, and
 does not have a closed form in general.
We present below a few choices of the degrees and the weight functions
that admit closed form solutions.
Concretely,
for $p=0$, it is known that LDDE  reduces to the standard Kernel Density Estimator (KDE) \cite{Loa96}:
\begin{eqnarray}
	\label{eq:p0}
	\hf_n(x) &=& \frac1n \sum_{i=1}^n  K\left(\frac{x-X_i}{h}\right) \Big/ \int K\left(\frac{u-x}{h}\right)\, du\;.
\end{eqnarray}
If we choose the step function  $K(u)=\ind(\|u\|\leq 1)$ with a local and data-dependent
choice of the bandwidth $h = \rho_{k,x}$ where $\rho_{k,x}$ is the $k$-NN distance from $x$, then
the above estimator recovers the popular $k$-NN density estimate as a special case, namely,
for $C_d=\pi^{d/2}/\Gamma(d/2+1)$,
\begin{eqnarray}
	\hf_n(x) &= & \frac{ \frac{1}{n} \sum_{i=1}^n \ind(\|X_i-x\|\leq \rho_{k,x})}{ {\rm Vol}\{u\in\reals^d : \|u-x\|\leq \rho_{k,x}\}}  \;=\; \frac{k}{n\, C_d \,\rho_{k,x}^d} \;.
	\label{eq:knn}
\end{eqnarray}
For higher degree local likelihood, we provide simple closed form solutions
and provide a proof in   Section~\ref{proof:prop}. Somewhat surprisingly, this result has eluded prior works \cite{lombardi2016nonparametric,vincent2003locally} and \cite{GVG15} which specifically attempted the evaluation for $p=2$. Part of the subtlety in the result is to critically use the fact that the parametric family  (eg., the polynomial family in \eqref{eq:localdensity}) need not be normalized themselves; the local log-likelihood maximization ensures that the resulting density estimate is correctly normalized so that it integrates to 1.
\begin{propo}{ \cite[Exercise 5.2]{loader2006local}}
	\label{pro:closedform}
	For a degree $p\in\{1,2\}$,
	the maximizer of local likelihood \eqref{eq:locallikelihood} admits a closed form solution,
	when using the Gaussian kernel $K(u)=e^{-\frac{\|u\|^2}{2}}$.
	In case of $p=1$,
\begin{eqnarray}
	\label{eq:p1}
	\hf_n(x) &=& \frac{S_0}{n(2\pi)^{d/2}h^{d}} \exp\left\{ -\frac{1}{2}\frac{1}{S_0^2}\|S_1\|^2 \right\}\;,
\end{eqnarray}
where $S_0\in\reals$ and $S_1\in\reals^d$ are defined for given $x\in\reals^d$ and $h\in\reals$ as
\begin{eqnarray}
	\label{eq:defS0}
	S_0 \equiv \sum_{j=1}^n e^{-\frac{\|X_j - x\|^2}{2h^2}  }\;,\; \;\;\;\;
	S_1 \equiv \sum_{j=1}^n  \frac{1}{h} (X_j-x)\, e^{-\frac{\|X_j - x\|^2}{2h^2}}\;. \;
\end{eqnarray}
In case of  $p=2$, for $S_0$ and $S_1$ defined as above,
\begin{eqnarray}
	\label{eq:p2}
	 \hf_n (x) &=& \frac{S_0}{n (2\pi)^{d/2} h^d |\Sigma|^{1/2} } \exp\Big\{ -\frac12 \frac{1}{S_0^2}S_1^T \Sigma^{-1}S_1  \Big\} \;,
\end{eqnarray}
where $|\Sigma|$ is the determinant and $S_2\in\reals^{d\times d }$ and $\Sigma\in\reals^{d\times d}$ are defined as
\begin{eqnarray}
	\label{eq:defS2}
	S_2 \equiv \sum_{j=1}^n  \frac{1}{h^2}(X_j-x)(X_j-x)^T\, e^{-\frac{\|X_j - x\|^2}{2h^2}}\;, \; \;\;\;\;
	\Sigma \equiv \frac{S_0S_2- S_1S_1^T}{S_0^2}\;,
\end{eqnarray}
where it follows from Cauchy-Schwarz that $\Sigma$ is positive semidefinite. 	
\end{propo}
One of the major drawbacks of
%state-of-the-art entropy and mutual information
the KDE and $k$-NN methods %\cite{Joe89,KL87,KSG04}
is the increased bias near the boundaries.
% A series of heuristics have been proposed recently in \cite{GVG14,GVG15} for $k$-NN estimators,
%by detecting boundaries and modifying the kernel to mitigate the bias.
LLDE provides a principled approach to automatically correct for the boundary bias, which takes effect only for $p\geq 2$ \cite{HJ96,She04}.
This explains the performance improvement for $p=2$ in the figure below (left panel),
and the gap increases with the correlation   as
boundary effect becomes more prominent.
We use the proposed estimators with $p\in\{0,1,2\}$ to estimate the mutual information between
two jointly Gaussian random variables with correlation $r$,
from $n=500$ samples, using resubstitution methods explained in the next sections. Each point is averaged over $100$ instances.

In the right panel, we generate i.i.d.\ samples from a 2-dimensional Gaussian with correlation 0.9, and
found local approximation $\hf(u-x^*)$ around $x^*$ denoted by the blue $*$ in the center.
Standard $k$-NN approach  fits a uniform distribution over a circle enclosing $k=20$ nearest neighbors (red circle).
The green lines are the contours of the degree-2 polynomial approximation with  bandwidth $h=\rho_{20,x}$.
 The figure illustrates that $k$-NN method suffers from boundary effect, where it underestimates the probability by over estimating the
 volume in \eqref{eq:knn}. However, degree-2 LDDE is able to correctly capture the local structure of the pdf,
 correcting for  boundary biases.

 Despite the advantages of the LLDE, it  requires the bandwidth to be data independent and vanishingly small (sublinearly in sample size) for consistency almost everywhere -- both of these are impediments to practical use since there is no obvious systematic way of choosing these hyperparameters.  On the other hand, if we restrict our focus to {\em functionals} of the density, then both these issues are resolved: this is the focus of the next section where we show that the bandwidth can be chosen to be based on {\em fixed} $k$-NN distances and the resulting universal bias easily corrected.

 \begin{figure}[h]
	\begin{center}
%	\put(-50,-3){correlation}
%	\put(-100,60){bias}
	\vspace{-.4cm}
	\includegraphics[width=.54\textwidth]{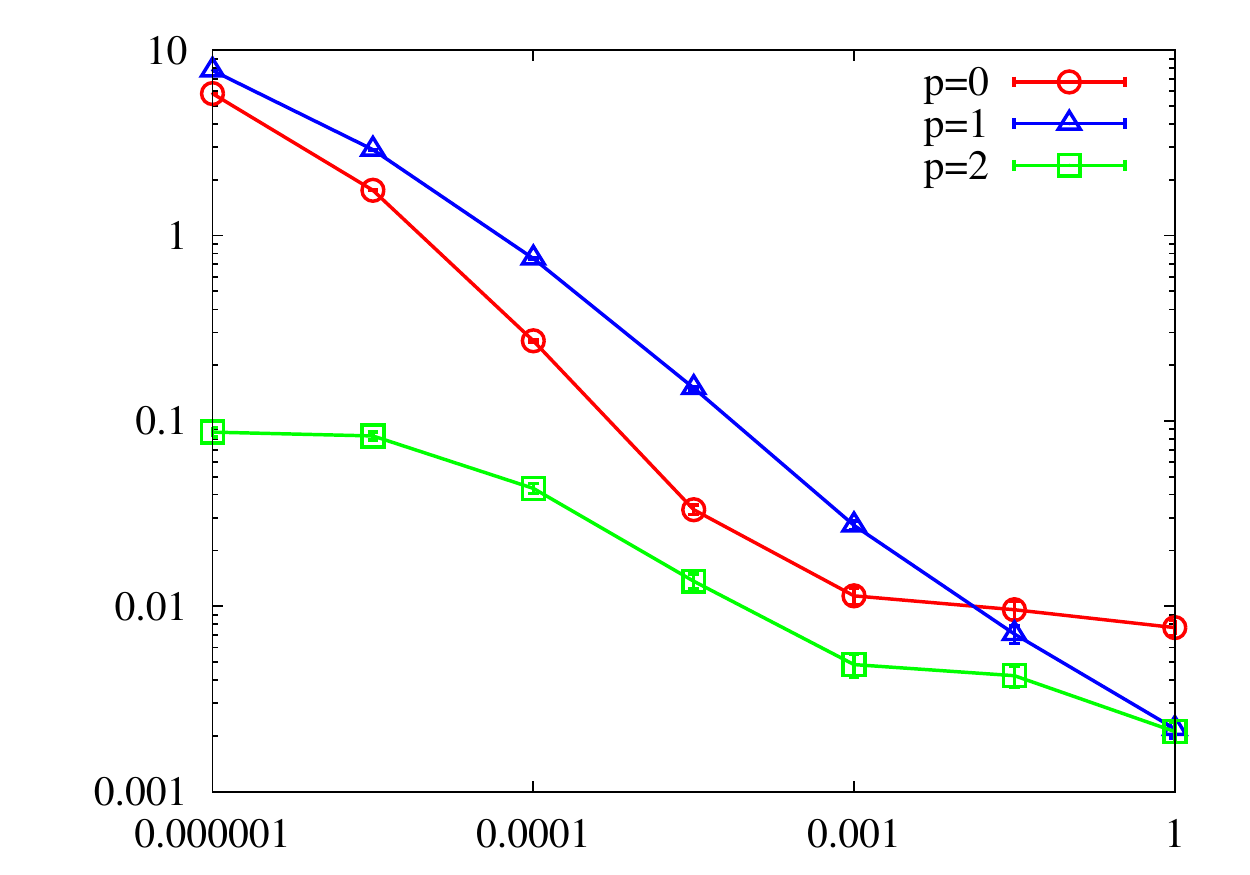}
	\put(-176,-7){$(1-r)$ where $r$ is correlation}
	\put(-266,100){$\E[(I-\widehat{I})^2]$}
	\includegraphics[width=.44\textwidth]{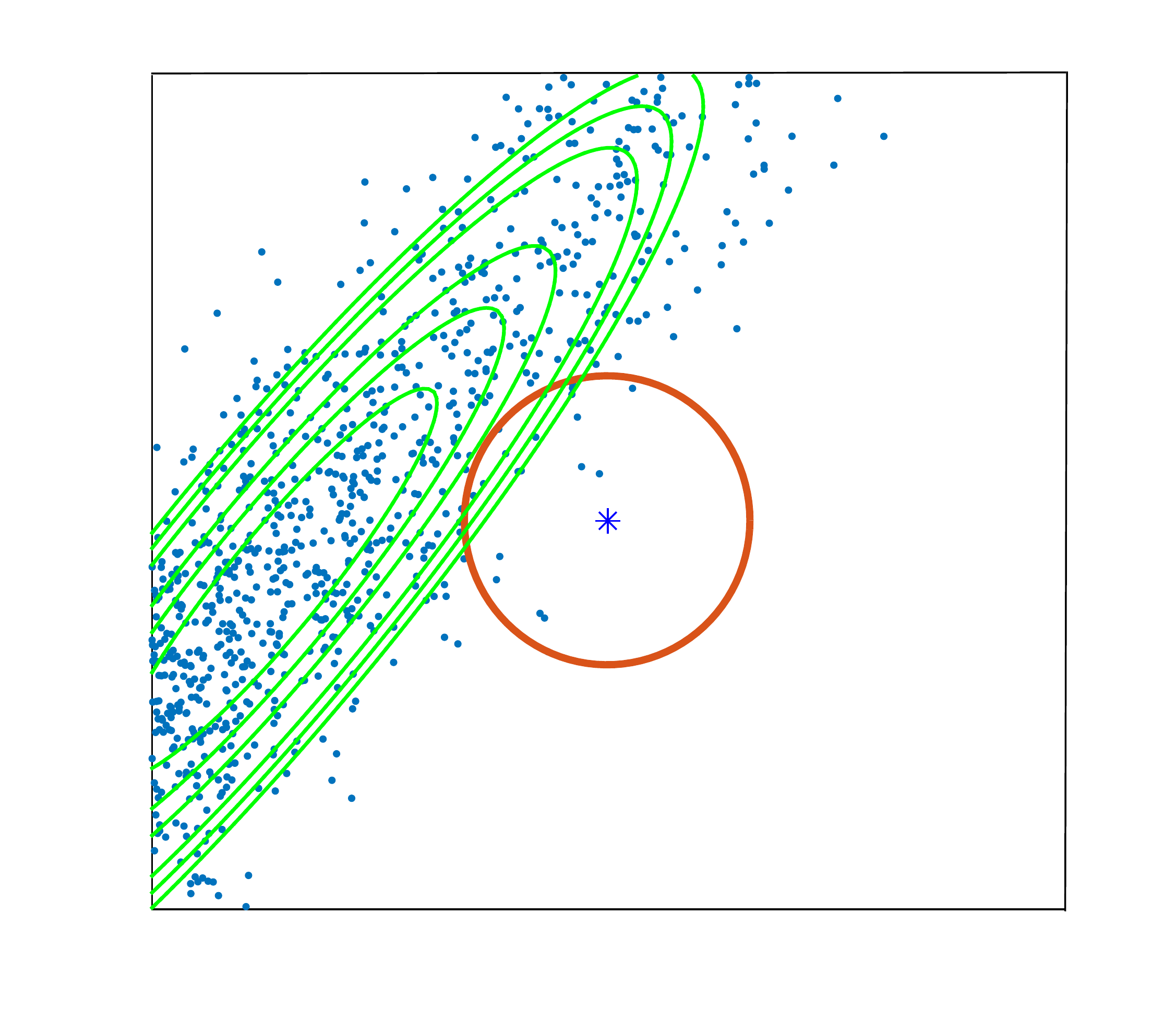}
	\put(-196,82){$X_1$}
	\put(-98,3){$X_2$}
	\end{center}
	\caption{
	The boundary bias becomes less significant and the gap closes as correlation decreases for
	estimating the mutual information (left).
	Local approximation around the blue $*$ in the center.
	The degree-2 local likelihood approximation (contours in green) automatically captures the local structure
	whereas the standard $k$-NN approach (uniform distribution in red circle) fails (left).
	}
	\label{fig:fit}
\end{figure}

\section{$k$-LNN Entropy Estimator }
\label{sec:main}
We consider {\em  resubstitution} entropy estimators of the form
$\hH(x) = -(1/n)\sum_{i=1}^n \log \hf_n(X_i)$ and  propose to use
the local likelihood density estimator in \eqref{eq:p2} and
  a  choice of bandwidth that is {\em local} (varying for each point $x$) and {\em adaptive} (based on the data).
Concretely, we choose, for each sample point $X_i$, the bandwidth $h_{X_i}$ to be the
the distance to its $k$-th nearest neighbor  $\rho_{k,i}$. Precisely,
we propose the following $k$-Local Nearest Neighbor ($k$-LNN) entropy estimator of degree-$2$:
\begin{eqnarray}
	\label{eq:resubstitute}
	\hH_{k{\rm LNN}}^{(n)}(X) &=& -\frac{1}{n}\sum_{i=1}^n  \left\{ \log \frac{S_{0,i}}{n(2\pi)^{d/2}\rho_{k,i}^d |\Sigma_i|^{1/2} }
	 - \frac12\frac{1}{S_{0,i}^2}  S_{1,i}^T  \Sigma_i^{-1} S_{1,i} \right\}  - B_{k,d}\;,
\end{eqnarray}
where subtracting $B_{k,d}$ defined in Theorem \ref{thm:unbiased} removes the asymptotic  bias,
and $k\in\Z^+$ is the only  hyper parameter determining the bandwidth. In practice $k$ is a small integer fixed to be in the range $4\sim 8$.
We
only use the $\lceil \log n \rceil$ nearest subset of samples  $\cS_{i}=\{j\in[n]\,:\, j\neq i \text{ and } \|X_i-X_j\|\leq  \rho_{\lceil\log n \rceil,i} \} $  in computing the quantities below:
\begin{eqnarray}
	S_{0,i} \equiv \sum_{j \in \cS_{i,m} } e^{-\frac{\|X_j - X_i\|^2}{2 \rho_{k,i}^2}  } \;, \;\;\;\;
	S_{1,i} \equiv \sum_{j \in \cS_{i,m} } \frac{1}{\rho_{k,i}} (X_j-X_i) e^{-\frac{\|X_j - X_i\|^2}{2 \rho_{k,i}^2}  }\;, \nonumber \\
	S_{2,i} \equiv  \sum_{j \in \cS_{i,m}  } \frac{1}{\rho_{k,i}^2} (X_j-X_i)(X_j-X_i)^T e^{-\frac{\|X_j - X_i\|^2}{2 \rho_{k,i}^2}  }\;,\;\;
	\Sigma_i \equiv \frac{S_{0,i}S_{2,i}- S_{1,i} S_{1,i}^T}{S_{0,i}^2}\;.
	\label{eq:defS}
\end{eqnarray}
The truncation is important %in practice
for computational efficiency, but the analysis works as long as $m=O(n^{{1/(2d)}-\varepsilon})$ for any positive $\varepsilon$ that can be arbitrarily small.  For a larger $m$, for example of $\Omega(n)$,
those neighbors that are further away have a different asymptotic behavior.
%and we discuss such phenomenon in Section \ref{sec:order}.
% (and als plays a role in the analysis   for  proving Theorem \ref{thm:unbiased}
%and also
We show in Theorem \ref{thm:unbiased} that the asymptotic bias   is {\em independent} of the underlying distribution and hence can be {\em precomputed} and removed,     under mild conditions  on  a twice continuously differentiable pdf $f(x)$ (cf.\ Lemma \ref{lem:order_stat} below). 
% Further, we provide an explicit formula characterizing $B_{k,d}$.
%We provide numerical evaluations in Table \ref{tbl:bias} via sampling and empirically approximating the expectation
%\eqref{eq:defBias}.
\begin{thm}
    \label{thm:unbiased}
%    Suppose $f$ is twice continuously differentiable.
    %has open support and the gradient $\|\nabla f\|$ and Hessian $\|H_f\|$ are bounded almost everywhere.
 For $k\geq 3$ and $X_1, X_2, \dots, X_n \in \mathbb{R}^d$ are i.i.d.\ samples from
 %   an underlying distribution $P$ with
   a twice continuously differentiable pdf   $f(x)$, then
    \begin{eqnarray}
    \lim_{n \to \infty} \E [\hH^{(n)}_{k{\rm LNN}}(X)  ]  &=&  H(X) \;,
    \end{eqnarray}
    where $B_{k,d}$ in \eqref{eq:resubstitute} is a  constant that only depends on $k$ and $d$. 
    Further, if $\E[ (\log f(X))^2]<\infty$ then 
     the variance of the proposed estimator is bounded by ${\rm Var} [\hH^{(n)}_{k{\rm LNN}}(X)  ]  =  O( (\log n)^2 /n)$.
\end{thm}
This proves the $L_1$ and $L_2$ consistency of the $k$-LNN estimator; we relegate the proof to Section~\ref{sec:proofmaintheorem}  for ease of reading the main part of the paper. 
The proof assumes Ansatz \ref{ansatz} (also stated in Section~\ref{sec:proofmaintheorem}, which states that  a certain exchange of limit holds.  
As noted in \cite{PPS10}, such an assumption is  common  in the literature on consistency of $k$-NN estimators, 
where it has been implicitly assumed 
in existing analyses of entropy estimators including \cite{KL87,GLMN05,LPS08,WKV09}, without explicitly stating that such assumptions are being made.  
%the supplementary material (where for compactness the univariate case is discussed in detail and which can be  generalized to higher dimensions $d$).
Our choice of a local adaptive bandwidth $h_{X_i}=\rho_{k,i}$
is crucial in ensuring that the asymptotic bias $B_{k,d}$
does not  depend on the underlying distribution $f(x)$.
This  relies on a fundamental connection to the theory of asymptotic order statistics made precise in Lemma \ref{lem:order_stat},
which also gives the explicit formula for the bias below.
%This gives the explicit formula for the bias.
%which can be evaluated numerically.
%For multivariate $X$, we conjecture that the asymptotic bias is still independent of $f(x)$, and give a formula for the bias in \eqref{eq:defBias} for general $d$.

The main idea is that the
empirical quantities used in the estimate \eqref{eq:defS}
converge in large $n$ limit to
similar quantities defined over order statistics.
We make this intuition precise in the next section.
We define  order statistics  over
i.i.d.\ standard exponential random variables $E_1,E_2,\ldots,E_m$  and
i.i.d.\ random variables $\xi_1,\xi_2,\ldots,\xi_m$ drawn uniformly (the Haar measure) over the unit sphere in $\reals^d$,
for a variable $m\in\Z^+$. We define for $\alpha\in\{0,1,2\}$,
\begin{eqnarray}
	\tilde{S}^{(m)}_{\alpha}  \equiv  \sum_{j=1}^{m} \xi_j^{(\alpha)}\,\frac{ (\sum_{\ell=1}^j E_\ell )^{\alpha}}{(\, \sum_{ \ell=1}^k E_\ell\,)^{\alpha}}
	\exp\left\{- \frac{(\,\sum_{\ell=1}^j E_\ell \,)^{2}}{2(\,\sum_{\ell=1}^k E_\ell\,)^{2}} \right\} \label{eq:S}\;,
\end{eqnarray}
where $\xi_j^{(0)}=1$, $\xi_j^{(1)}=\xi_j\in\reals^d$, and $\xi_j^{(2)}=\xi_j \xi_j^T\in\reals^{d\times d}$, and let
$\tS_\alpha = \lim_{m\to \infty} \tS_\alpha^{(m)}$ and $\tSigma = (1/\tS_0)^2 (\tS_0\tS_2 -\tS_1 \tS_1^T)$.
 We show that the limiting $\tS_\alpha$'s are well-defined (in the proof of Theorem \ref{thm:unbiased}) and are directly related to the bias terms in the resubstitution
 estimator of entropy:
% When $d=1$, $\xi_j$'s are  Rademacher random variables.
\begin{eqnarray}
	B_{k,d} = \E[\, \log ( \sum_{\ell=1}^k E_\ell) + \frac{d}{2} \log  2\pi -\log C_d - \log \tS_0 + \frac12 \log  \big|\tSigma \big|  +  (\frac{1}{2\tS_0^2} \tS_1^T \tSigma^{-1}\tS_1 )  \,] \;.
	\label{eq:defBias}
\end{eqnarray}

In practice, we propose using a fixed small $k$ such as five.
For $k\leq3$ the estimator has a very large variance, and
numerical evaluation of the corresponding bias also converges slowly.
For some typical choices of $k$, we provide approximate evaluations below, where
$0.0183(\pm 6)$ indicates empirical mean $\mu=183\times10^{-4}$ with confidence interval   $6\times10^{-4}$.
In these numerical evaluations, we truncated the summation at $m=50,000$.
Although we prove that $B_{k,d}$ converges in $m$,
in practice, one can choose $m$  based on the number of samples and $B_{k,d}$ can be evaluated for that $m$.

%\subsection{}
{\bf Theoretical contribution}:
Our key technical innovation is
 a fundamental connection
 between nearest neighbor statistics and asymptotic order statistics, stated below as  Lemma \ref{lem:order_stat}:
we show that the (normalized) distances $\rho_{\ell,i}$'s jointly converge to the standardized uniform  order statistics and
the directions $(X_{j_\ell} -X_i)/\|X_{j_\ell} -X_i\|$'s converge to independent uniform distribution (Haar measure) over the unit sphere.

\begin{table}[h]
\begin{center}
  \begin{tabular}{  c c | c | c | c | c | c | c   |}
	\cline{3-8}
    & & \multicolumn{6}{|c|}{$k$} \\ \cline{3-8}
     & &  $4$ & $5$ & $6$ & $7$ & $8$ & $9$ \\ \cline{3-8}   \hline
     \multicolumn{1}{ |c  }{\multirow{2}{*}{$d$} } &
     \multicolumn{1}{ |c|| }{$1$} & -0.0183($\pm$6) & -0.0233($\pm$6) & -0.0220($\pm$4)  & -0.0200($\pm$4) & -0.0181($\pm$4) &  -0.0171($\pm$3)
     % for k=10, -0.0157(3), for k=11, -0.0144(3)
         \\ \cline{2-8}
     \multicolumn{1}{ |c  }{} &
     \multicolumn{1}{ |c|| }{$2$} & -0.1023($\pm$5) & -0.0765($\pm$4) & -0.0628($\pm$4) & -0.0528($\pm$3) & -0.0448($\pm$3)&    -0.0401($\pm$3)  \\ \cline{1-8}
  \end{tabular}
\end{center}
     \caption{Numerical evaluation of $B_{k,d}$, via sampling
       $1,000,000$ instances  for each pair $(k,d)$.}
     \label{tbl:bias}
	\end{table}
	
Conditioned on $X_i=x$,
the proposed estimator uses nearest neighbor statistics on
$Z_{\ell,i} \equiv X_{j_\ell} - x $ where
$X_{j_\ell}$ is the $\ell$-th nearest neighbor from $ x$ such that
$ Z_{\ell,i} = ((X_{j_\ell} -X_i)/\|X_{j_\ell} -X_i\|) \rho_{\ell,i}$.
Naturally, all the  techniques we develop in this paper generalize to any estimators that depend on
the nearest neighbor statistics $\{Z_{\ell,i}\}_{i,\ell\in[n]}$ -- and the value of such a general result is demonstrated later (in Section~\ref{sec:general}) when we evaluate the bias in similarly inspired entropy estimators \cite{GVG14,GVG15,lombardi2016nonparametric,KL87}. % from the literature.

% Consider the joint distribution of $(Z_{1,i},\ldots,Z_{m_n,i})$.
%Fix $X_i = x$. We will show that $\lim_{n \to \infty} \E[H_i | X_i = x] = \log f(x) + C_k$ for some constant $C_k$. Notice that $H_i$ is a function of random variables $ \rho_{k,i}$ and $S_{0,i}, S_{1,i}, S_{2,i}$. Furthermore, $S_{0,i}, S_{1,i}, S_{2,i}$ are functions of $\{\rho_{j,i}\}_{j=1}^{m_n}$. The following lemma gives the distribution of $\{\rho_{j,i}\}_{j=1}^{m_n}$ as $n$ goes to infinity.
\begin{lemma}
    \label{lem:order_stat}
    Let $E_1, E_2, \dots, E_m$ be i.i.d.\ standard exponential random variables and
    $\xi_1 , \xi_2, \dots, \xi_m$ be i.i.d.\ random variables  drawn uniformly over the unit $(d-1)$-dimensional sphere
    in $d$ dimensions,
    independent of the $E_i$'s.
    Suppose $f$ is twice continuously differentiable and $x \in \mathbb{R}^d$ satisfies that there exists $\varepsilon > 0$ such that $f(a) > 0$, $\|\nabla f(a)\| = O(1)$ and $\|H_f(a)\| = O(1)$ for any $\|a - x\| < \varepsilon$.
    Then for any $m = O( \log n)$,
    we have the following convergence  conditioned on  $X_i = x$:
    \begin{eqnarray}
    \lim_{n \to \infty} d_{\rm TV} ( (c_d nf(x))^{1/d} (\, Z_{1,i},  \dots, Z_{m,i} \,) \;,\;
    (\, \xi_1 E_1^{1/d}, \dots , \xi_{m}(\sum_{\ell=1}^{m} E_\ell)^{1/d} \,) ) = 0 \;.
    \end{eqnarray}
    where $d_{\rm TV}(\cdot,\cdot)$ is the total variation and $c_d$ is the volume of unit Euclidean ball in $\mathbb{R}^d$.%$\|X - Y\|_{TV} = \sup_B |\,\Pr\{X \in B\} - \Pr\{Y \in B\}\,|$ is the total variation distance of two random variables $X$ and $Y$.
\end{lemma}

%Given this, ...

\begin{figure}[h]
	\begin{center}
	\includegraphics[width=.45\textwidth]{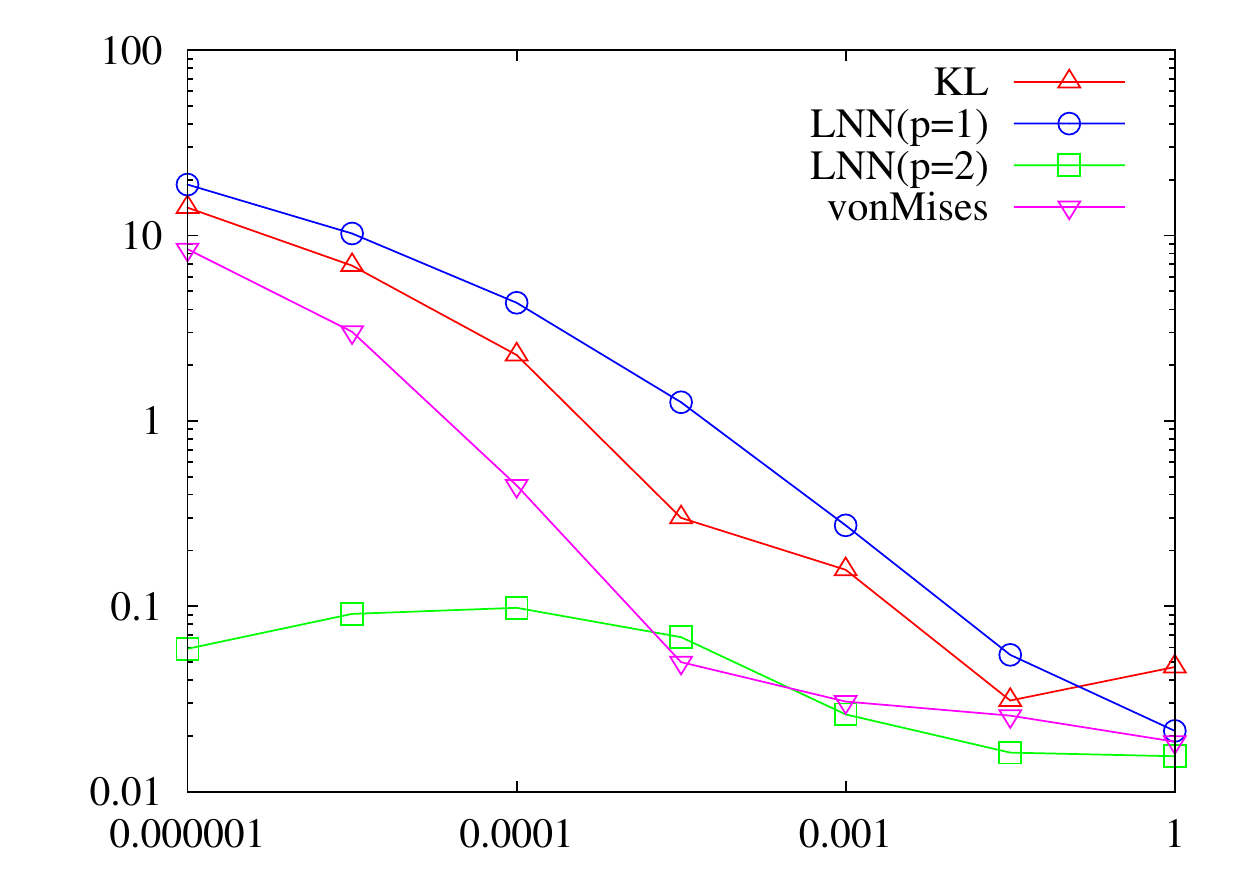}
	\put(-247,83){$ \E[(\hH-H)^2] $}
	\put(-165,-10){ $(1-r)$ where $r$ is correlation}
	\includegraphics[width=.45\textwidth]{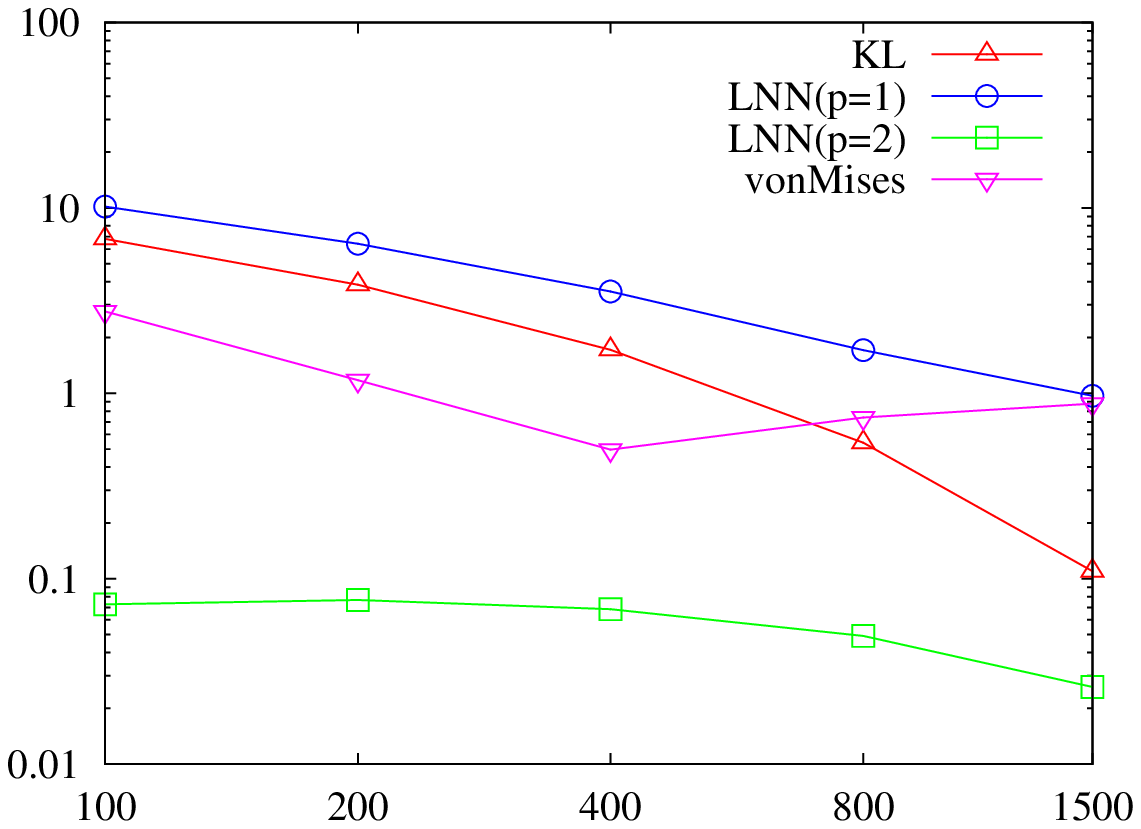}
	\put(-147,-10){number of samples $n$}
	\end{center}
	\caption{Degree-2 $k$-LNN  outperforms other state-of-the-art estimators for entropy estimation. }
	\label{fig:entropy}
\end{figure}

{\bf Empirical contribution}:
 Numerical experiments suggest that the proposed estimator
outperforms
state-of-the-art entropy estimators,
and the gap increases with correlation.
The idea of using $k$-NN distance as bandwidth
for entropy estimation was originally proposed
by Kozachenko and Leonenko in \cite{KL87}, and is a special case of the $k$-LNN method we propose with degree $0$ and a step kernel.
We refer to Section \ref{sec:general} for a formal comparison.
Another popular resubstitution entropy estimator is to
use  KDE in \eqref{eq:p0} %and resubstituting it to estimate entropy
 \cite{Joe89}, which
 is a special case of the  $k$-LNN method with degree $0$, and
the Gaussian kernel is used in simulations.
As comparison, we also study a new estimator \cite{KKPW15} based on von Mises expansion (as opposed to simple re-substitution)
 which has an improved convergence rate in the large sample regime.
In Figure \ref{fig:entropy} (left), we draw $100$ samples i.i.d.\ from two standard Gaussian random variables with correlation $r$, and plot resulting  mean squared error averaged over $100$ instances.
The ground truth, in this case is  $H(X) = \log(2 \pi e) + 0.5 \log(1-r^2)$.
% N = 500, truncation = 80
On the right, we repeat the same simulation for fixed $r=0.99999$ and varying number of samples and $m=7\log_e n$.
%The main reason we use Gaussian distributions is the convenience in computing the ground truths, and
%for more extensive experimental results, we refer to the journal version of this paper \cite{GOV16journal}.

\begin{figure}[h]
	\begin{center}
	\includegraphics[width=.45\textwidth]{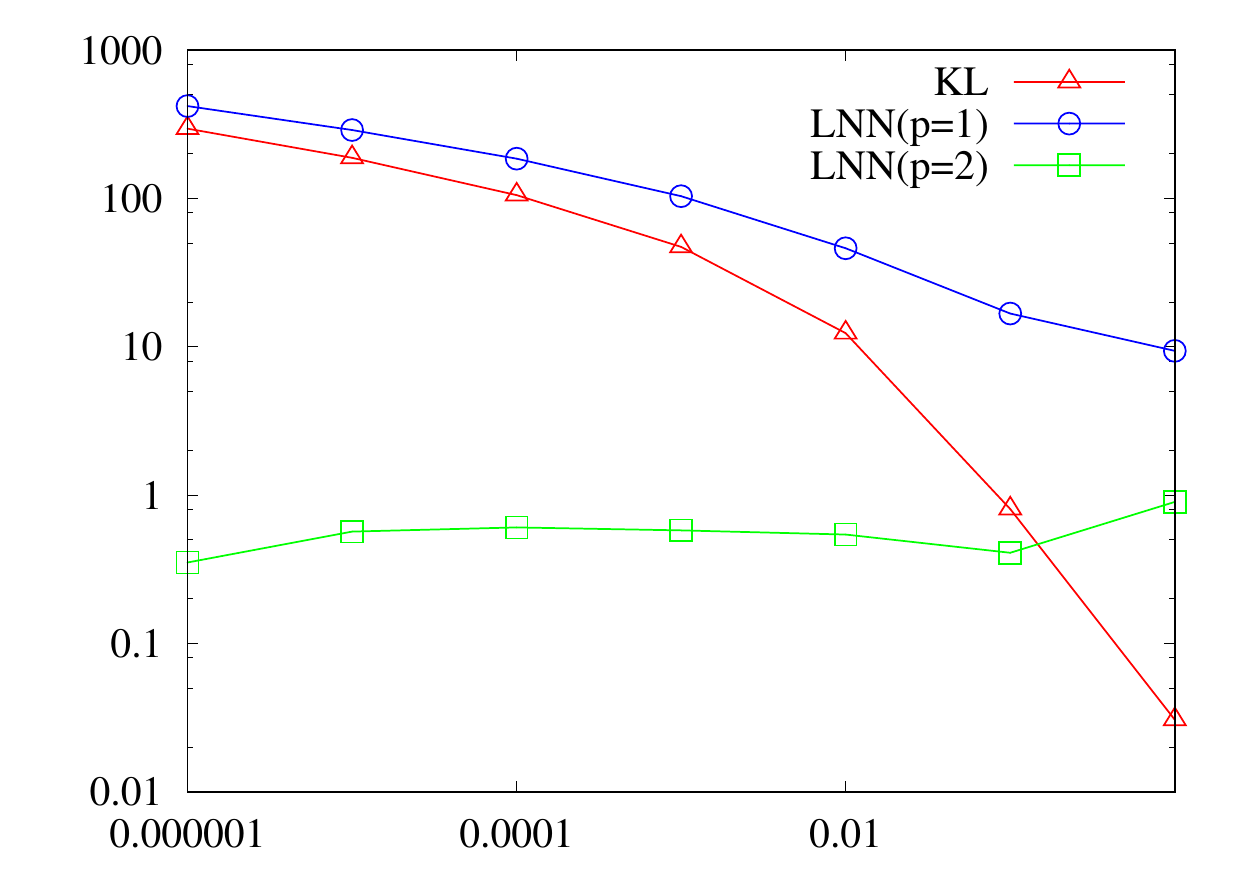}
	\put(-247,83){$ \E[(\hH-H)^2] $}
	\put(-165,-10){ $(1-r)$ where $r$ is correlation}
	\includegraphics[width=.45\textwidth]{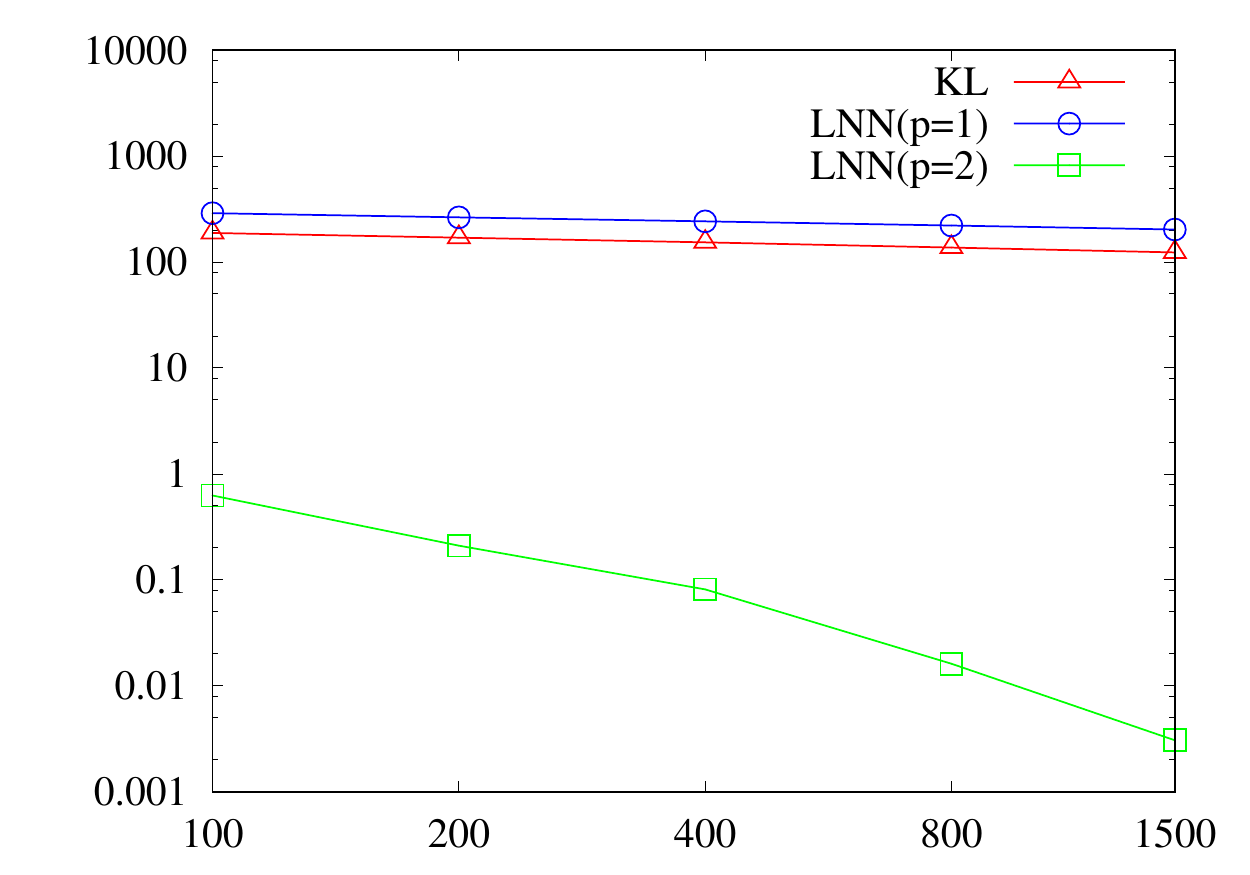}
	\put(-147,-10){number of samples $n$}
	\end{center}
	\caption{Degree-2 $k$-LNN  outperforms other state-of-the-art estimators for high-dimensional entropy estimation. }
	\label{fig:entropy_6d}
\end{figure}

In Figure \ref{fig:entropy_6d}, we repeat the same simulation for 6 standard Gaussian random variables with ${\rm Cov}(X_1, X_2) = {\rm Cov}(X_3, X_4) = {\rm Cov}(X_5, X_6) = r$ and ${\rm Cov}(X_i, X_j) = 0$ for other pairs $(i,j)$. On the left, we draw $100$ i.i.d. samples with various $r$. We plot resulting mean squared error averaged over $100$ instances.
The ground truth is  $H(X) = 3\log(2 \pi e) + 1.5 \log(1-r^2)$.
% N = 500, truncation = 80
On the right, we repeat the same simulation for fixed $r=0.99999$ and varying number of samples and $m=7\log_e n$.

\begin{figure}[h]
	\begin{center}
	\includegraphics[width=.45\textwidth]{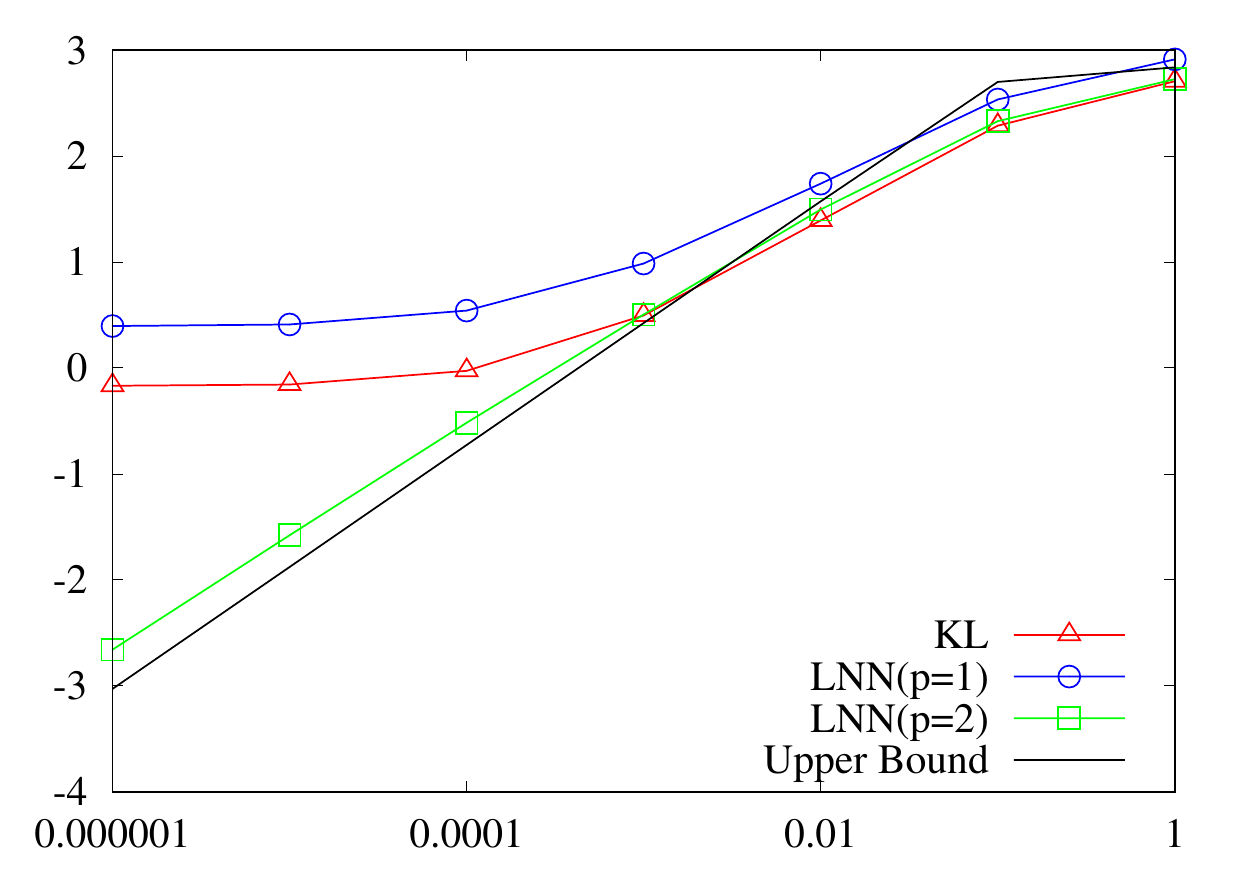}
	\put(-230,83){$ \E[ \hH ] $}
	\put(-165,-10){ $(1-r)$ where $r$ is correlation}
	\includegraphics[width=.45\textwidth]{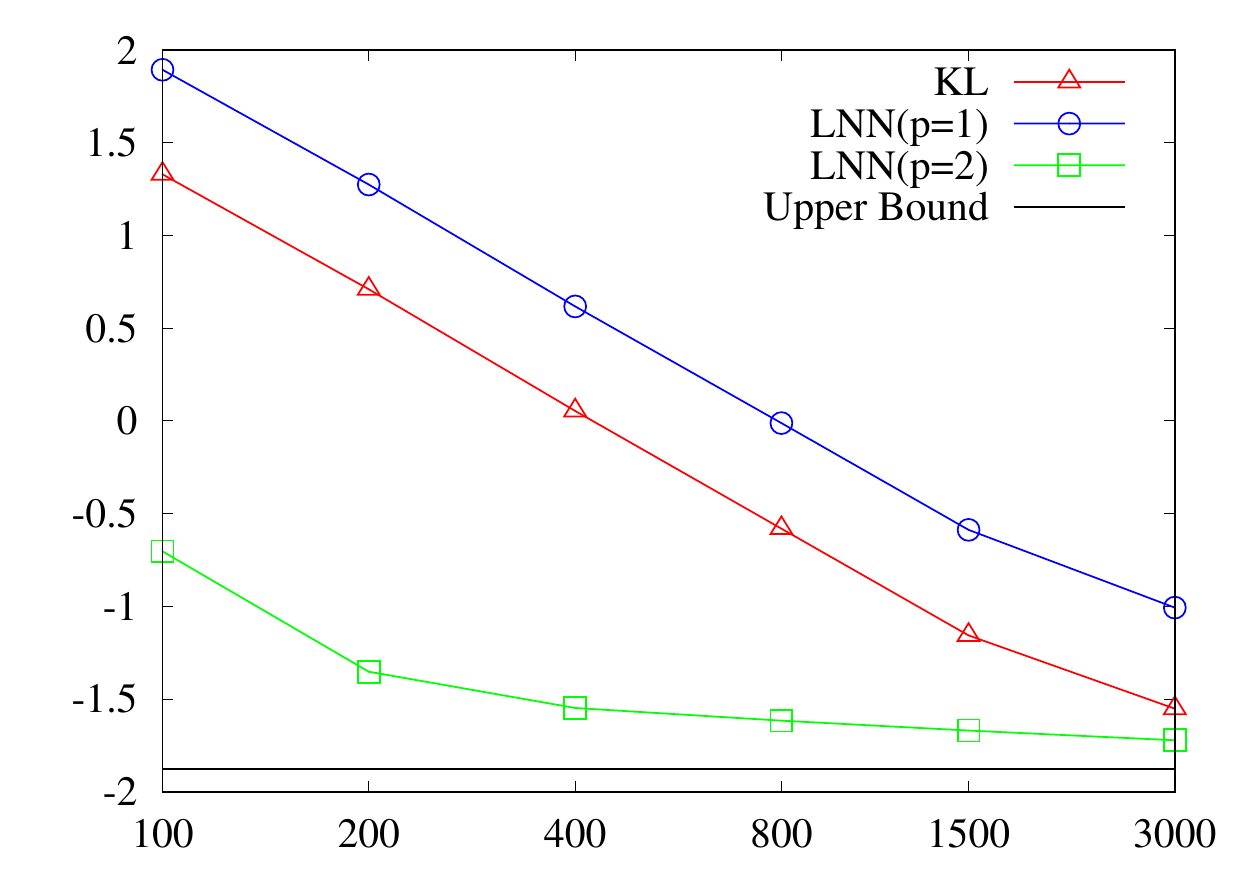}
	\put(-147,-10){number of samples $n$}
	\end{center}
	\caption{Degree-2 $k$-LNN  outperforms other state-of-the-art estimators for non-Gaussian entropy estimation. }
	\label{fig:mixture}
\end{figure}

In Figure \ref{fig:mixture} (left), we draw $100$ samples i.i.d.\ from a mixture of two joint Gaussian distributions with zero mean and covariance $\begin{pmatrix} 1& r \\ r & 1 \end{pmatrix}$ and $\begin{pmatrix} 1& -r \\ -r & 1 \end{pmatrix}$, respectively, and plot resulting average estimate over $100$ instances.
Here we plot an upper bound of the ground truth $H(X) \leq \log(2) + \log(2 \pi e) + 0.5 \log(1-r^2)$ for $r \geq 0.9$.
On the right, we repeat the same simulation for fixed $r=0.99999$ and varying number of samples and $m=7\log_e n$.

% ---------------------------------------------------------------------------------------------------------------------------------
\section{Universality of the  $k$-LNN approach }
\label{sec:general}
{In this section, we show that Theorem \ref{thm:unbiased} holds universally for a general family of entropy estimators,
 specified by the choice of
$k\in\Z^+$, degree $p\in\Z^+$,
and a kernel $K:\reals^d\to\reals$, thus allowing a unified view of several seemingly disparate entropy estimators \cite{KL87,GVG14,GVG15,lombardi2016nonparametric}.  The template of the entropy estimator is the following:
given $n$ i.i.d.\ samples, we first compute the local density estimate by
 maximizing the local likelihood \eqref{eq:locallikelihood}
with bandwidth $\rho_{k,i}$, and then
resubstitute it to estimate entropy: $\hH_{k,p,K }^{(n)}(X) = -(1/n)\sum_{i=1}^n \log \hf_n(X_i) $.}
\begin{thm}
	\label{thm:unify}
	For  the family of estimators described above, % with any choice of a degree $p\in \Z^+$ and a  kernel  $K:\reals^d\to\reals$,
	under the hypotheses of Theorem~\ref{thm:unbiased},
	if the solution to the maximization $ \ha(x) = \arg\max_a \cL_x(f_{a,x})$ exists for all $x\in\{X_1,\ldots,X_n\}$, then
	for any choice of $k\geq p+1$, $p\in\Z^+$, and $K:\reals^d\to\reals$,
	the asymptotic bias is independent of the underlying distribution:
	\begin{eqnarray}
		\lim_{n\to\infty} \E[\hH^{(n)}_{k,p,K}(X) ] &=& H(X) + \tB_{k,p,K,d}
		\label{eq:unify}\;,
	\end{eqnarray}
	for some constant $\tB_{k,d,p,K}$ that only depends on $k,p,K$ and $d$.
	% the dimension $d$, and the choice of degree $p$, 	weight function $K$, and the hyper parameter $k$ that determines the bandwidth.
\end{thm}
We provide a proof in Section~\ref{sec:proof_unify}.
Although   in general there is no simple analytical characterization  of the asymptotic bias $\tB_{k,p,K,d}$  it can be readily numerically computed:
 since $\tB_{k,p,K,d}$ is independent of the underlying distribution,
one can run the estimator over i.i.d.\ samples from {\em any} distribution and numerically approximate the bias for {\em any} choice of the parameters.
However, when the maximization  $ \ha(x) = \arg\max_a \cL_x(f_{a,x})$ admits a closed form solution,
as is the case with proposed $k$-LNN, then $\widetilde{B}_{k,p,K,d}$ can be characterized explicitly in terms of uniform order statistics.

This family of estimators is general:   for instance,
the popular  KL estimator
 is a special case with $p=0$ and a step kernel $K(u)=\ind(\|u\|\leq1)$.
 \cite{KL87} showed (in a remarkable result at the time) that the asymptotic bias is independent of the dimension $d$
and can be  computed exactly to be
% and was removed from the estimate:
%\begin{eqnarray}
%	\hH_{\rm KL}(X)
%	&=&  - \frac1n \sum_{i=1}^n \log\frac{k}{n C_d \rho_{k,i}^d} - \tB_{k,0,{\rm step},d} \;,
%\end{eqnarray}
%where $\tB_{k,0,{\rm step},d}=
$\log n - \psi(n) + \psi(k)-\log k$ and $\psi(k)$ is the digamma function defined as $\psi(x) = \Gamma^{-1}(x)d\Gamma(x)/dx$.
%Further, not only is this bias  independent of the underlying distribution, but also independent of the dimension $d$.
The dimension independent nature of this asymptotic bias term (of $O(n^{-1/2})$ for $d=1$ in \cite[Theorem 1]{TV96} and $O(n^{-1/d})$ for general $d$ in \cite{GOV16})
is special to the choice of $p=0$ and the step kernel;
we explain this in detail in Section \ref{sec:proof_unify}, later in the paper.
Analogously, the estimator in \cite{GVG14} can be viewed as a special case with $p=0$ and an ellipsoidal step kernel.
%Together with the numerical evaluations in Table \ref{tbl:bias},  this suggests that ... [I am waiting for the table to be filled.]

% ---------------------------------------------------------------------------------------------------------------------------------
\section{$k$-LNN Mutual information estimator}

Given an entropy estimator  $\hH_{\rm KL}$, mutual information can be estimated:
$\hI_{\rm 3KL} = \hH_{\rm KL}(X)+\hH_{\rm KL}(Y)-\hH_{\rm KL}(X,Y)$.
In  \cite{KSG04}, Kraskov and St\"ogbauer and Grassberger
 introduced $\hI_{\rm KSG}(X;Y)$ by coupling the choices of the bandwidths.
The joint entropy is estimated in the usual way,
but for the marginal entropy, instead of using $k$NN distances from $\{X_j\}$,  the bandwidth
$h_{X_i}=\rho_{k,i}(X,Y)$ is chosen, which is the $k$ nearest neighbor distance from $(X_i,Y_i)$ for the joint data $\{(X_j,Y_j)\}$.
Consider
$\hI_{\rm 3LNN}(X;Y) = \hH_{k{\rm LNN}}(X)+\hH_{k{\rm LNN}}(Y)-\hH_{k{\rm LNN}}(X,Y)$.
Inspired by \cite{KSG04}, we introduce the following novel mutual information estimator we denote by
$\hI_{\rm LNN-KSG}(X;Y)$. % = \hH_{(\rho_k(X,Y))-{\rm LNN}}(X) + \hH_{(\rho_k(X,Y))-{\rm LNN}}(Y) - \hH_{k{\rm LNN}}(X,Y) $,
where for the joint $(X,Y)$ we use the LNN entropy estimator we proposed in \eqref{eq:resubstitute},
and for the marginal entropy we use the bandwidth $h_{X_i}=\rho_{k,i}(X,Y)$ coupled to the joint estimator.
%$ \hH_{(\rho_k(X,Y))-{\rm LNN}}(X)$ uses is the $k$ nearest neighbor distance from $(X_i,Y_i)$ for the joint data $\{(X_j,Y_j)\}$.
Empirically, we observe  $\hI_{\rm KSG}$ outperforms $\hI_{\rm 3KL}$ everywhere, validating the use of correlated bandwidths.
However, the performance of $\hI_{{\rm LNN-KSG}}$ is similar to $\hI_{3{\rm LNN}}$--sometimes better and sometimes worse.

\begin{figure}[h]
	\begin{center}
	\includegraphics[width=.45\textwidth]{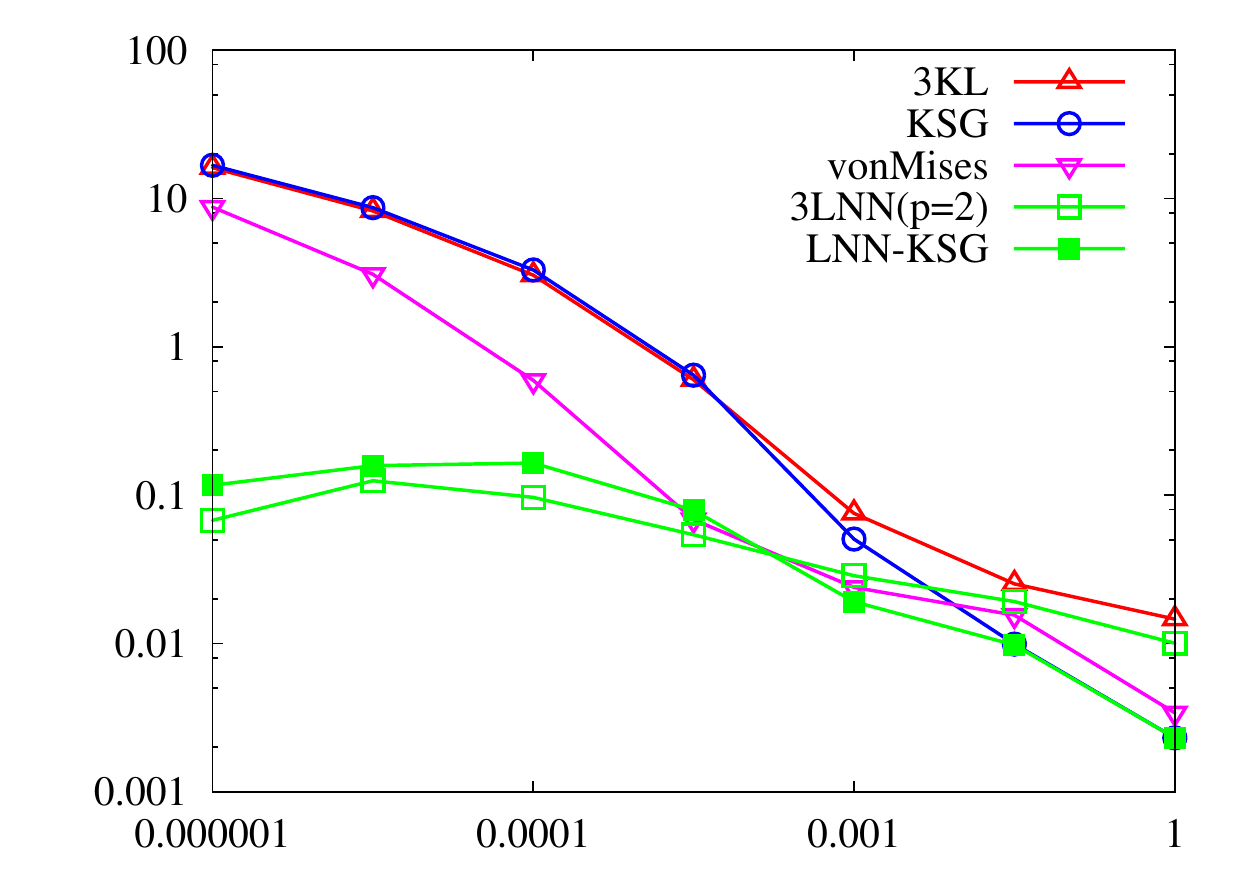}
	\put(-230,90){\small$ \E[(\hI-I)^2] $}
	\put(-159,-10){ $(1-r)$ where $r$ is correlation}
	\hspace{1.1 cm}
	\includegraphics[width=.45\textwidth]{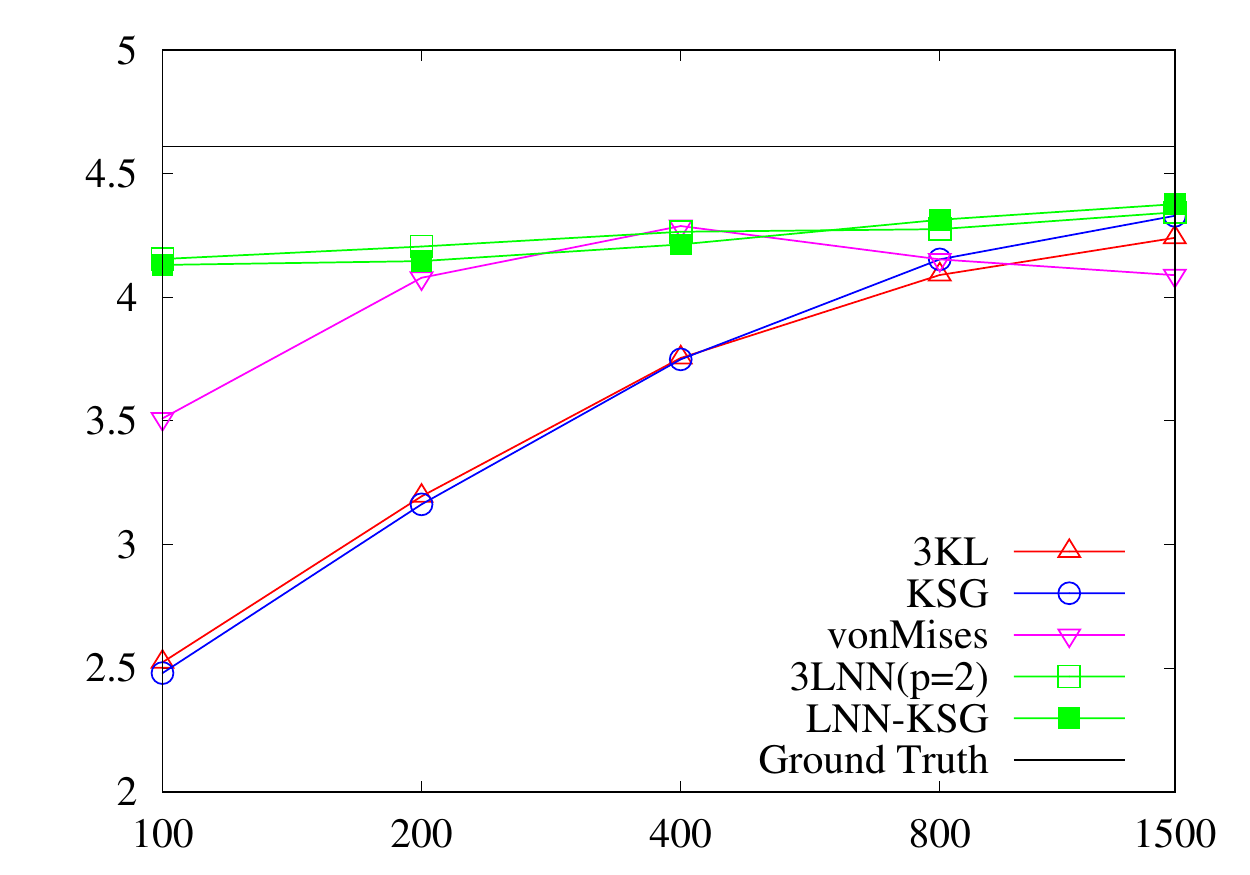}
	\put(-238,90){\small$ \E[\hI(X;Y)] $}
	\put(-139,-10){ number of samples $n$}
	\end{center}
	\caption{Proposed $\hI_{\rm LNN-KSG}$ and $\hI_{\rm 3LNN}$ outperform other state-of-the-art  estimators.}
	\label{fig:mi}
\end{figure}

In Figure \ref{fig:mi} (left), we estimate mutual information under the same setting as in Figure \ref{fig:entropy} (left).
For most regimes of correlation $r$,  both 3LNN and  LNN-KSG outperforms other state-of-the-art estimators.
The gap increases with correlation $r$.
 On the right, we draw i.i.d. samples from two random variables $X$ and $Y$, where
 $X$ is uniform over $[0,1]$ and $Y=X+U$, where $U$ is uniform over $[0,0.01]$ independent of $X$.
In the large sample limit, all estimators find the correct mutual information.
The plot show how sensitive the estimates are, in the small sample regime.
Both LNN and LNN-KSG are significantly more robust compared to other approaches.
 Mutual information estimators have been recently proposed in \cite{GVG14,GVG15,lombardi2016nonparametric} based on   local likelihood maximization.
However,  they involve heuristic choices of hyper-parameters or solving elaborate optimization and numerical integrations,
which are far from being easy to implement.

%We are replicating the same setting as \cite{GVG15}.
%Although an implementation of the local Gaussian approximation algorithm introduced in \cite{GVG15} is not available,
%the performance should be comparable to LNN-KSG

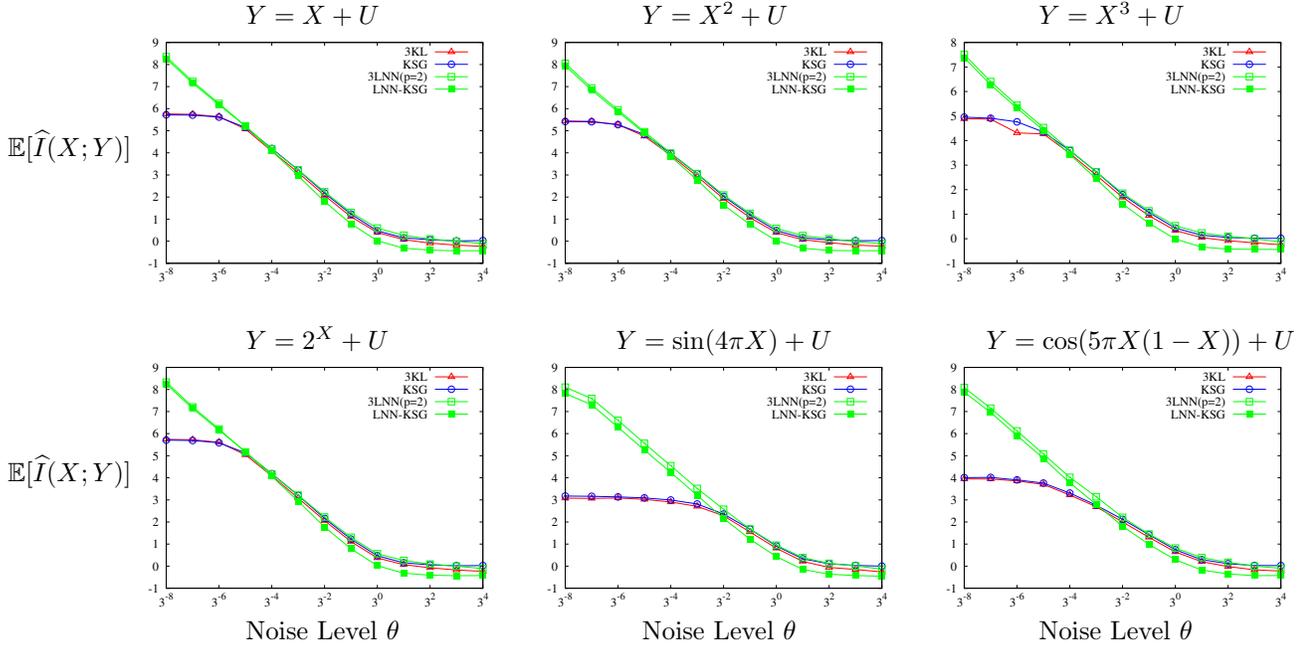
\begin{figure}[h]
    \begin{center}$
        \begin{array}{lll}
        \includegraphics[width=.3\textwidth]{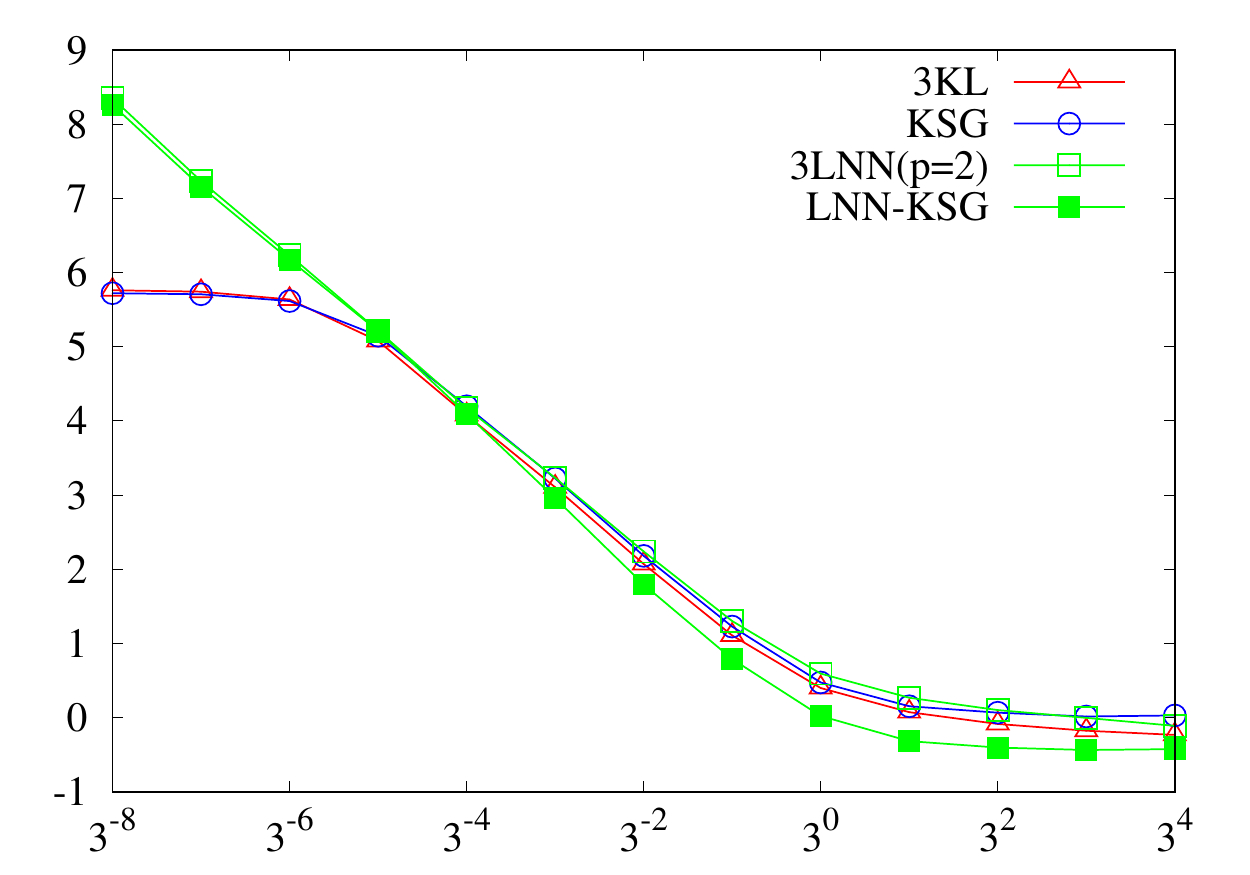}&
        \includegraphics[width=.3\textwidth]{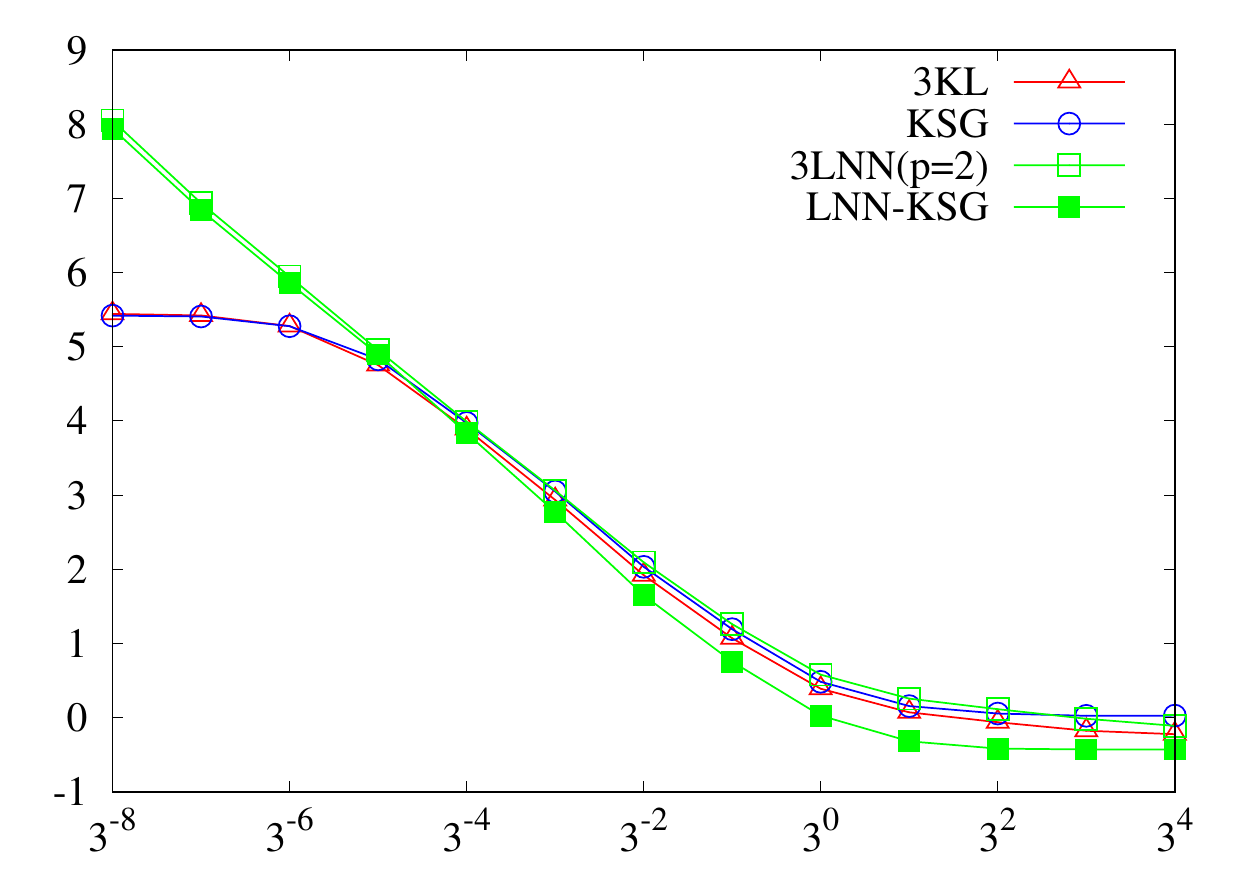}&
        \includegraphics[width=.3\textwidth]{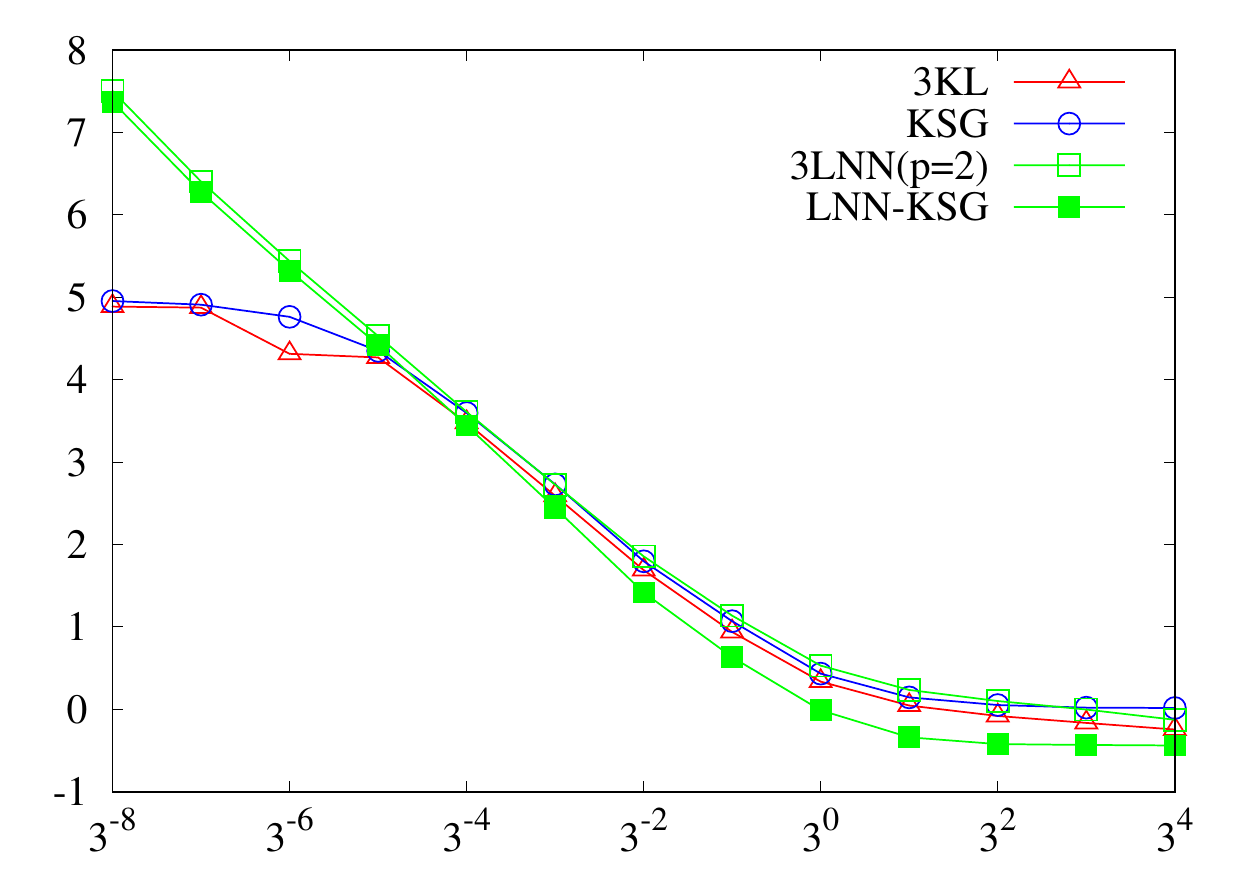}
        \put(-490,50){$\E [\hI(X;Y)]$}
        \put(-400,100){$Y = X+U$}
        \put(-250,100){$Y = X^2+U$}
        \put(-100,100){$Y = X^3+U$}
        \end{array}$
    \end{center}
    \begin{center}$
        \begin{array}{lll}
        \includegraphics[width=.3\textwidth]{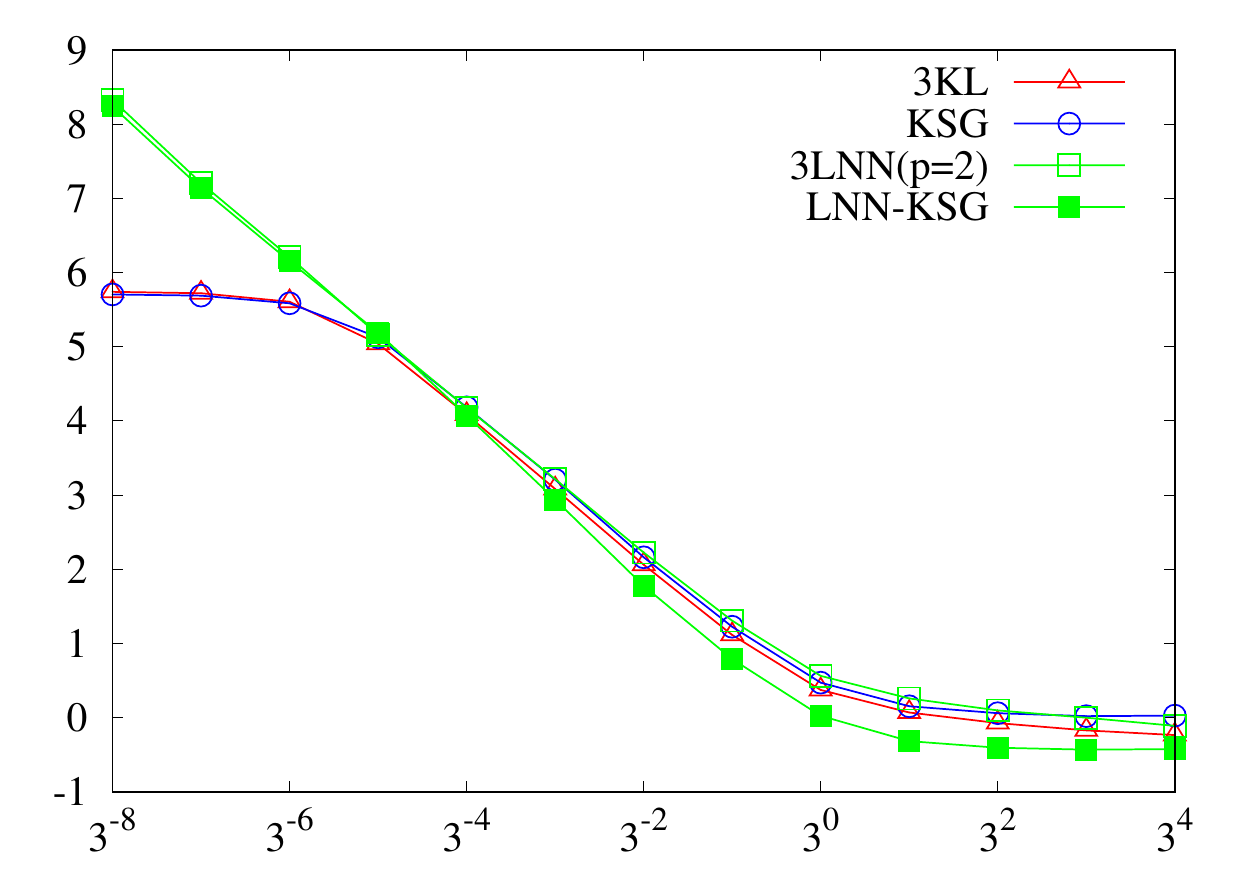}&
        \includegraphics[width=.3\textwidth]{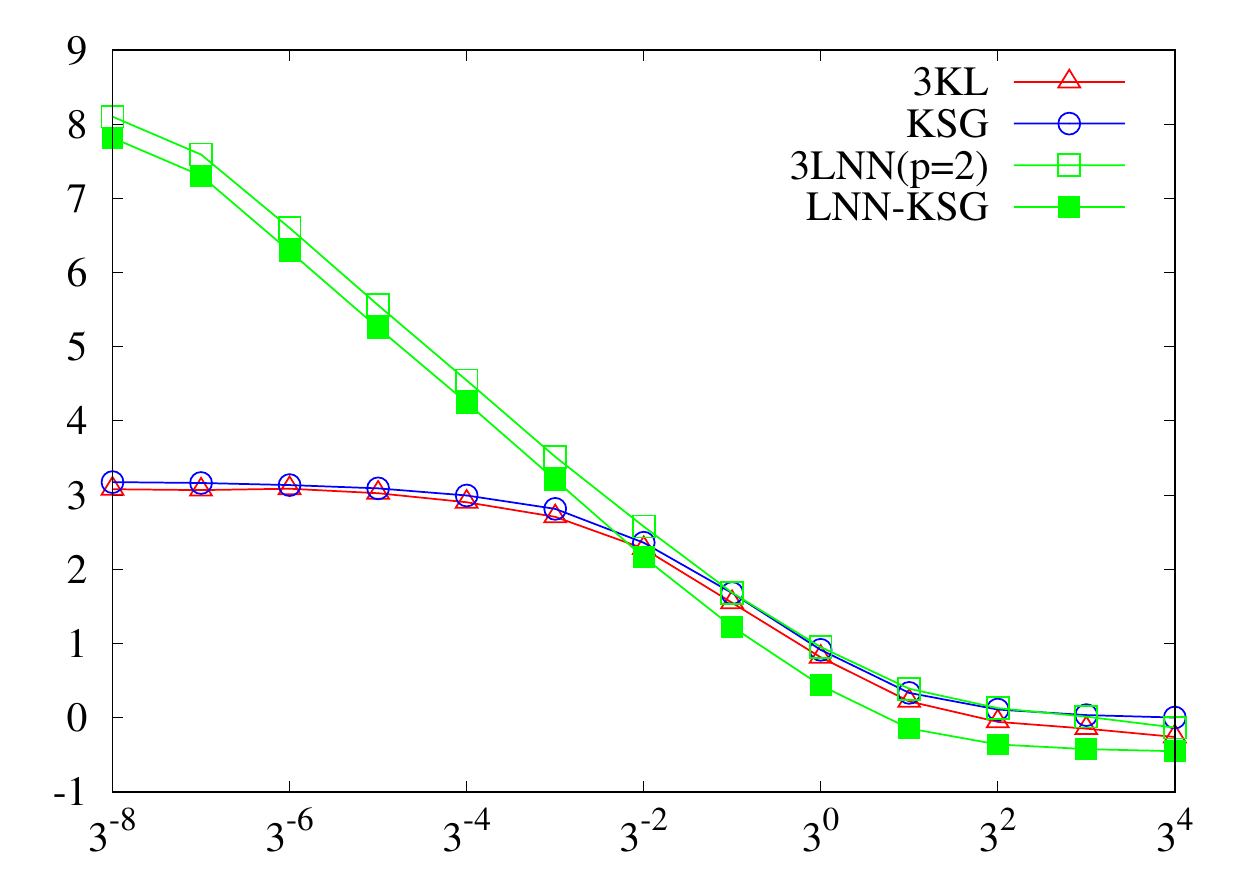}&
        \includegraphics[width=.3\textwidth]{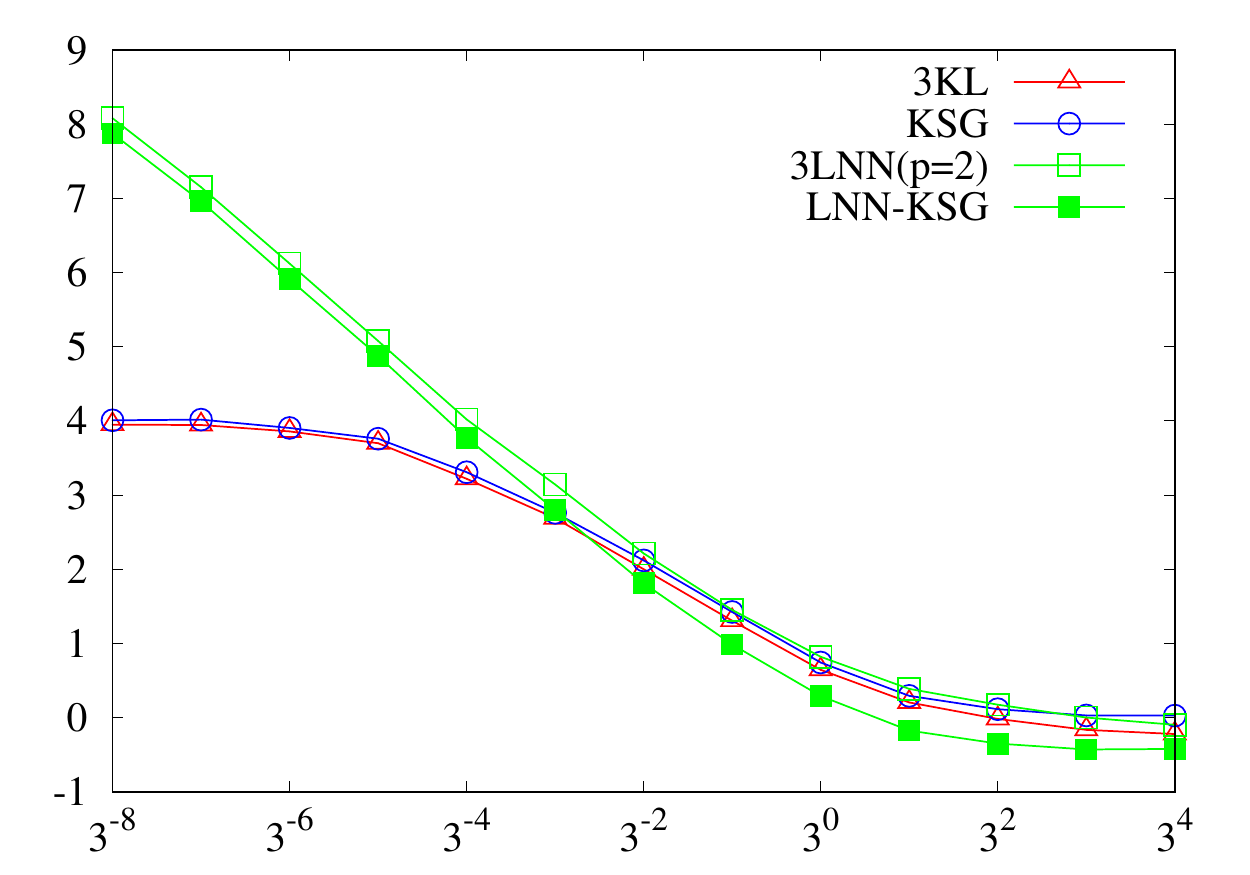}
        \put(-490,50){$\E [\hI(X;Y)]$}
        \put(-400,100){$Y = 2^X+U$}
        \put(-260,100){$Y = \sin(4\pi X)+U$}
        \put(-120,100){$Y = \cos(5\pi X(1-X))+U$}
        \put(-400,-10){Noise Level $\theta$}
        \put(-250,-10){Noise Level $\theta$}
        \put(-100,-10){Noise Level $\theta$}
    \end{array}$
    \end{center}
    \caption{Functional relationship test for mutual information estimators. Proposed $\hI_{\rm LNN-KSG}$ and $\hI_{\rm 3LNN}$ outperform other state-of-the-art  estimators.}
	\label{fig:gsg_1}
\end{figure}

In Figure~\ref{fig:gsg_1}, we test the mutual information estimators for $Y = f(X) + U$, where $X$ is uniformly distributed over $[0,1]$ and $U$ is uniformly distributed over $[0, \theta]$, independent of $X$, for some noise level $\theta$. Similar simulation were studied in~\cite{GVG15}. We draw 2500 i.i.d. sample points for each relationship.
The plot show that for small noise level $\theta$, i.e., near-functional related random variables, our proposed estimators $\hI_{3LNN}$ and $\hI_{LNN-KSG}$ perform much better than 3KL and KSG estimators. Also our proposed estimators can handle both linear and nonlinear functional relationships.

\begin{figure}[h]
    \begin{center}$
        \begin{array}{ll}
        \includegraphics[width=.4\textwidth]{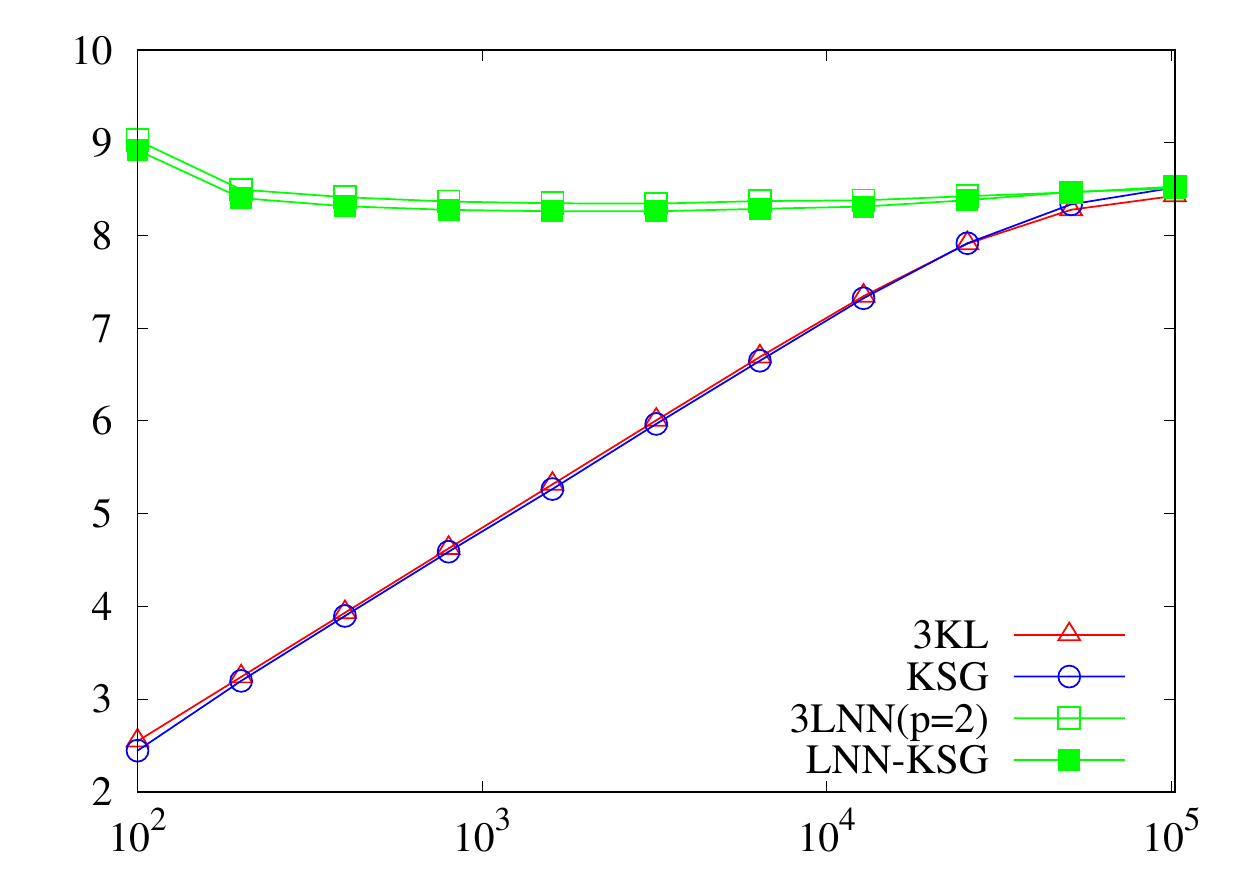}&
        \includegraphics[width=.4\textwidth]{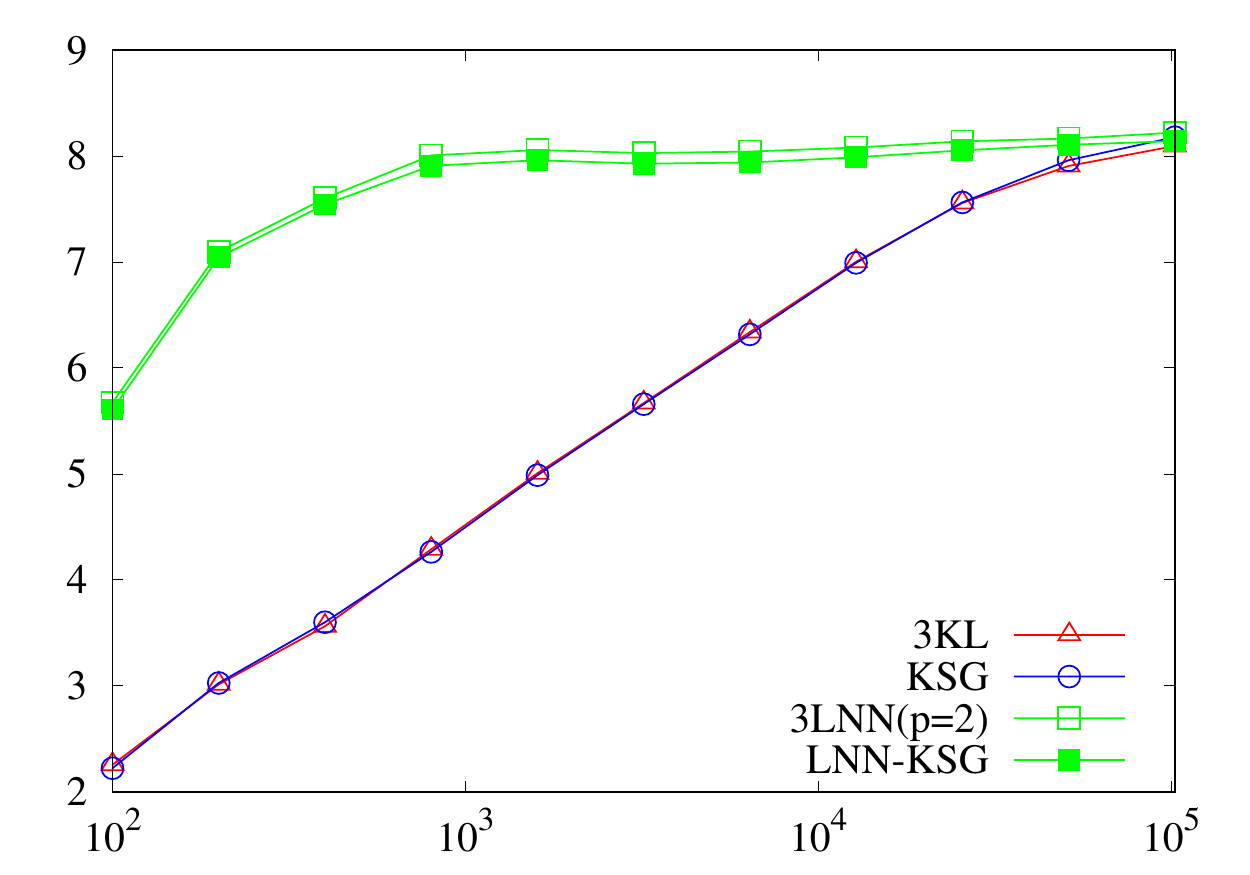}
        \put(-425,70){$\E [\hI(X;Y)]$}
        \put(-320,130){$Y = X_1+U$}
        \put(-120,130){$Y = X_1^2+U$}
        \end{array}$
    \end{center}
    \begin{center}$
        \begin{array}{ll}
        \includegraphics[width=.4\textwidth]{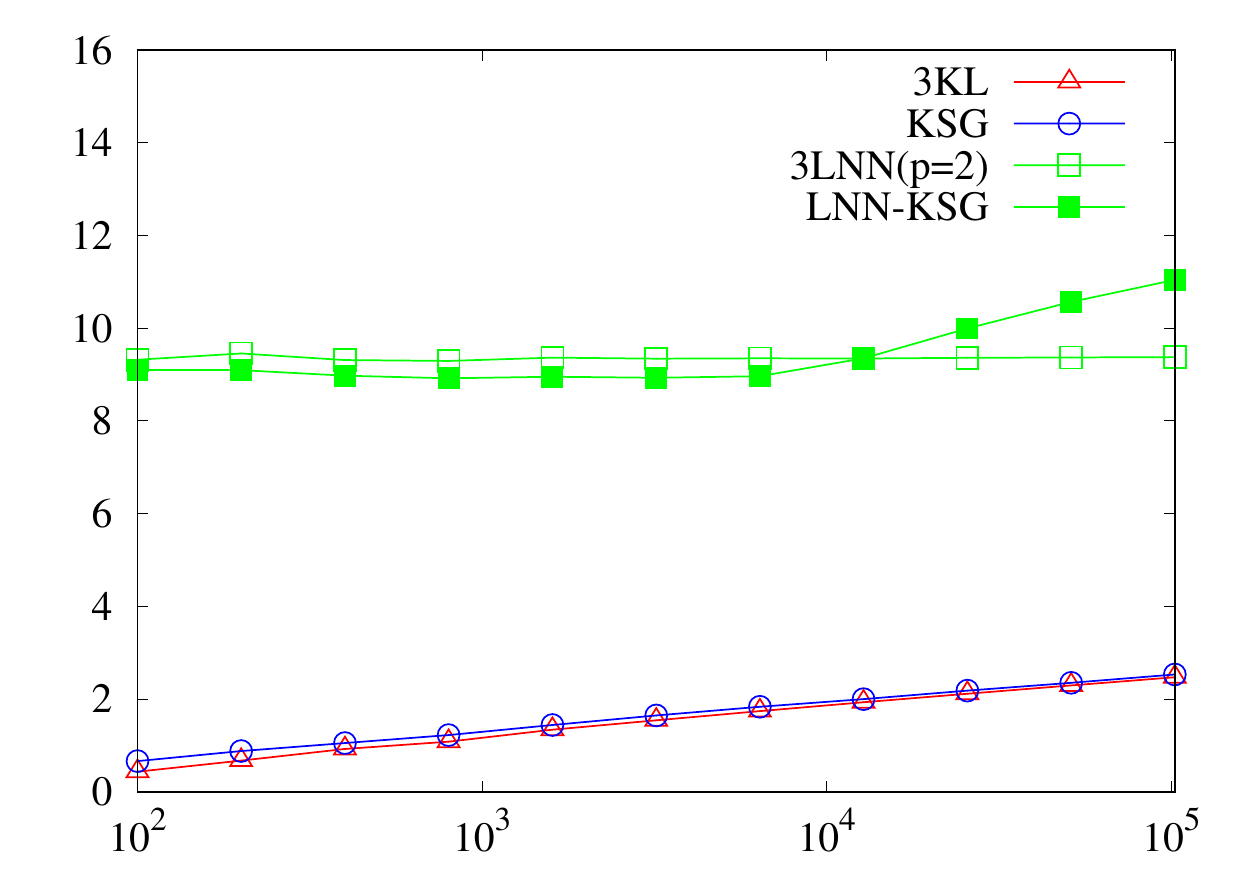}&
        \includegraphics[width=.4\textwidth]{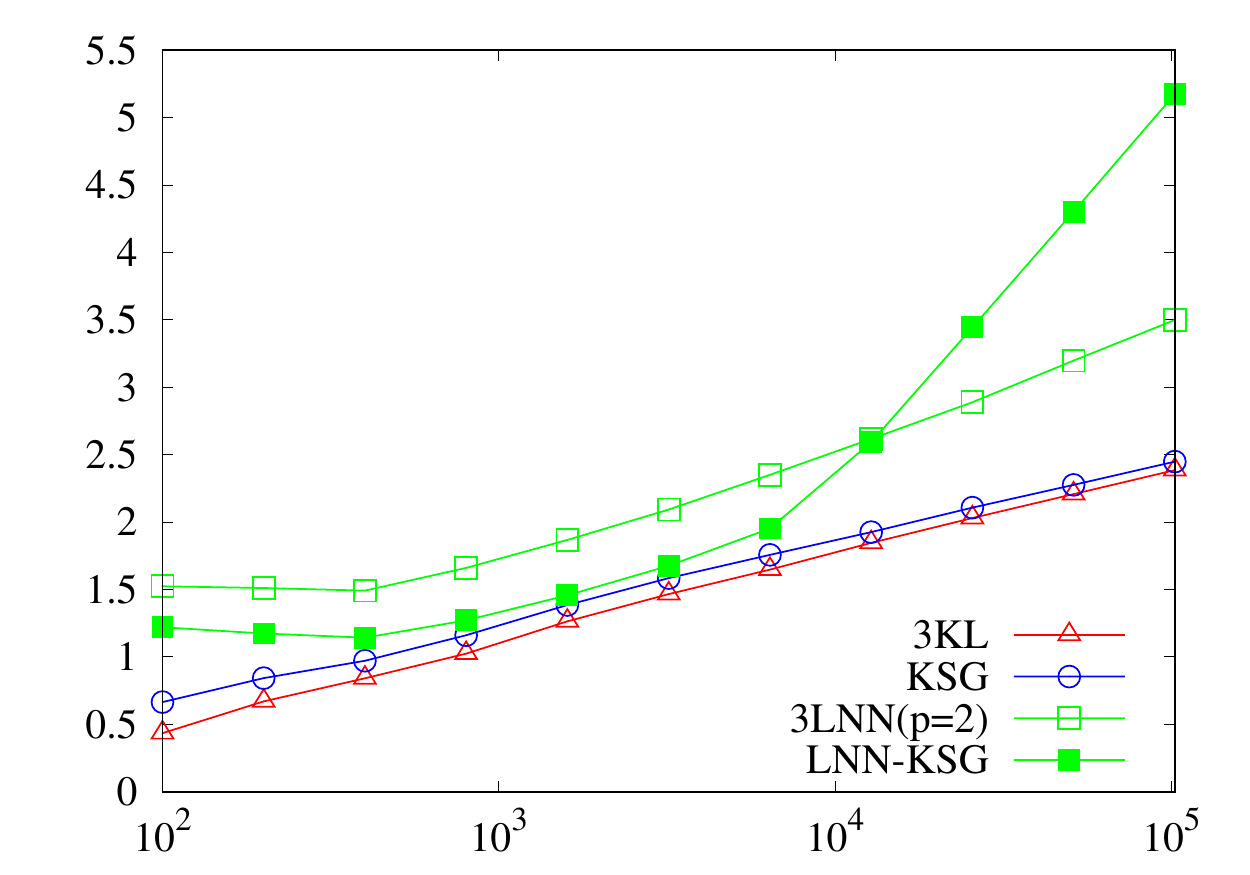}
        \put(-425,70){$\E [\hI(X;Y)]$}
        \put(-350,130){$Y = X_1+X_2+X_3+X_4+U$}
        \put(-150,130){$Y = X_1^2+X_2^2+X_3^2+X_4^2+U$}
        \put(-330,-10){number of samples}
        \put(-130,-10){number of samples}
    \end{array}$
    \end{center}
    \caption{Estimated Mutual Information of low/high-dimensional relationships}
	\label{fig:gsg_2}
\end{figure}

In Figure~\ref{fig:gsg_2}, we test our estimators on linear and nonlinear relationships for both low-dimensional  ($D=2$) and high-dimensional ($D=5$). Here $X_i$'s are uniformly distributed over $[0,1]$ and $U$ is uniformly distributed over $[-3^8/2,3^8/2]$, independently of $X_i$'s. Similar simulation were studied in~\cite{GVG14}. We can see that our estimators $\hI_{3LNN}$ and $\hI_{LNN-KSG}$ converges much faster than $\hat{I}_{3KL}$ and $\hI_{KSG}$.

\section{Breaking the bandwidth barrier}
\label{sec:bandwidth}

While $k$-NN distance based bandwidth are routine in practical usage  \cite{She04}, the main finding of this work is that they also turn out to be the ``correct" mathematical choice for the purpose of
asymptotically unbiased estimation of
 an integral functional such as the entropy: $- \int f(x)\log f(x)$; we briefly discuss the ramifications below.
%We start the discussion with  LLDE  and  KDE methods,
%leading to the $k$NN method in the celebrated work by Kozachenko and Leonenko \cite{KL87} and its  connection to  order statistics.
Traditionally, when the goal is to estimate  $f(x)$, it is well known that the bandwidth should
%decreasing sub linearly with $n$ but not too fast.
satisfy  $h\to0$ and $nh^d\to\infty$, for
KDEs  to be consistent. % under mild assumptions.
%For $d=1$, a sharper analysis  proposes
%using $h = O(n^{-1/5} )$ to achieve the optimal tradeoff between
%bias $O(h^{2}+(nh)^{-1})$ and variance $O((nh)^{-1})$.
%both  LLDE and KDE  are    consistent under mild assumptions.
%When $d=1$, a sharper analysis in \cite{HJ96} proposes
%using $h = O(n^{-1/(2p+5)} )$ to achieve the optimal tradeoff between
%bias $O(h^{p+2}+(nh)^{-1})$ and variance $O((nh)^{-1})$ for degree-$p$ LLDE.
 As a rule of thumb,  $h= 1.06 \widehat{\sigma} n^{-1/5}$ is suggested when $d=1$
 where $\widehat{\sigma}$ is the sample standard deviation \cite[Chapter 6.3]{Was06}.
%In practice, however, data dependent choices using
%cross-validation and plug-in methods
% give faster convergence rates.
%
%However in practice, instead of fixing the bandwidth for all $x$,
%a more flexible choice of $h_x$ that
%depends locally on $x$ is preferred
%\cite{Jon90}.
On the other hand, when estimating entropy, as well as other integral functionals,
it is known  that resubstitution estimators of the form $ -(1/n) \sum_{i=1}^n \log \hf(X_i)$
achieve variances scaling as $O(1/n)$ independent of the bandwidth
% as long as $h=\Omega(n^{-1/d})$
\cite{LWL12}.
% This is due to the inherent averaging effect of the estimators: every sample contributes to the estimate regardless of the bandwidth.
%Notice that $h = \Omega(n^{-1/d})$ is necessary. Otherwise we have vanishing neighboring samples within distance $h$ of any $x$.
This allows for a bandwidth as small as $O(n^{-1/d})$. % for entropy estimation.

\begin{figure}[h]
	\begin{center}
	\includegraphics[width=.45\textwidth]{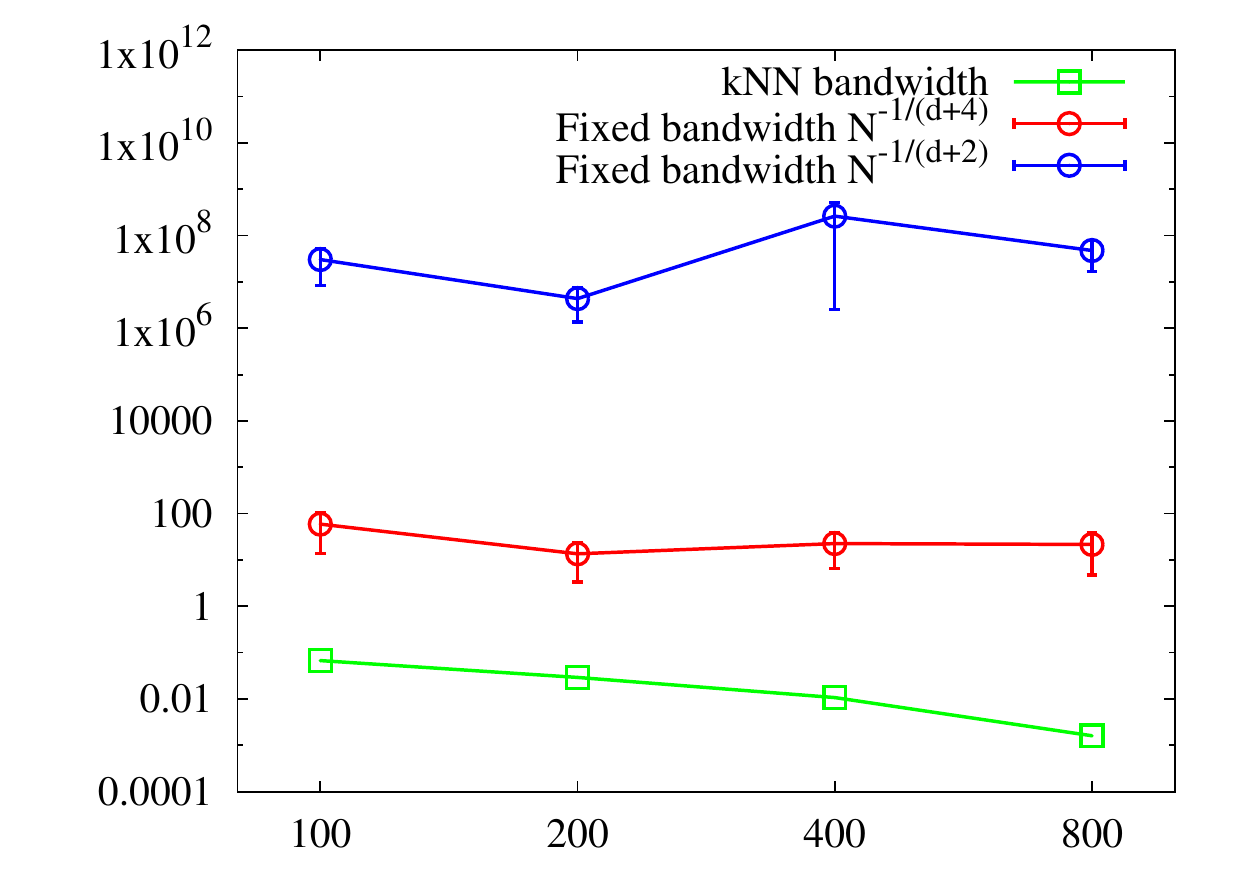}
	\put(-139,-10){number of samples $n$}
	\put(-240,98){\small$ \E[(\hI-I)^2] $}
	\hspace{1.cm}
	\includegraphics[width=.45\textwidth]{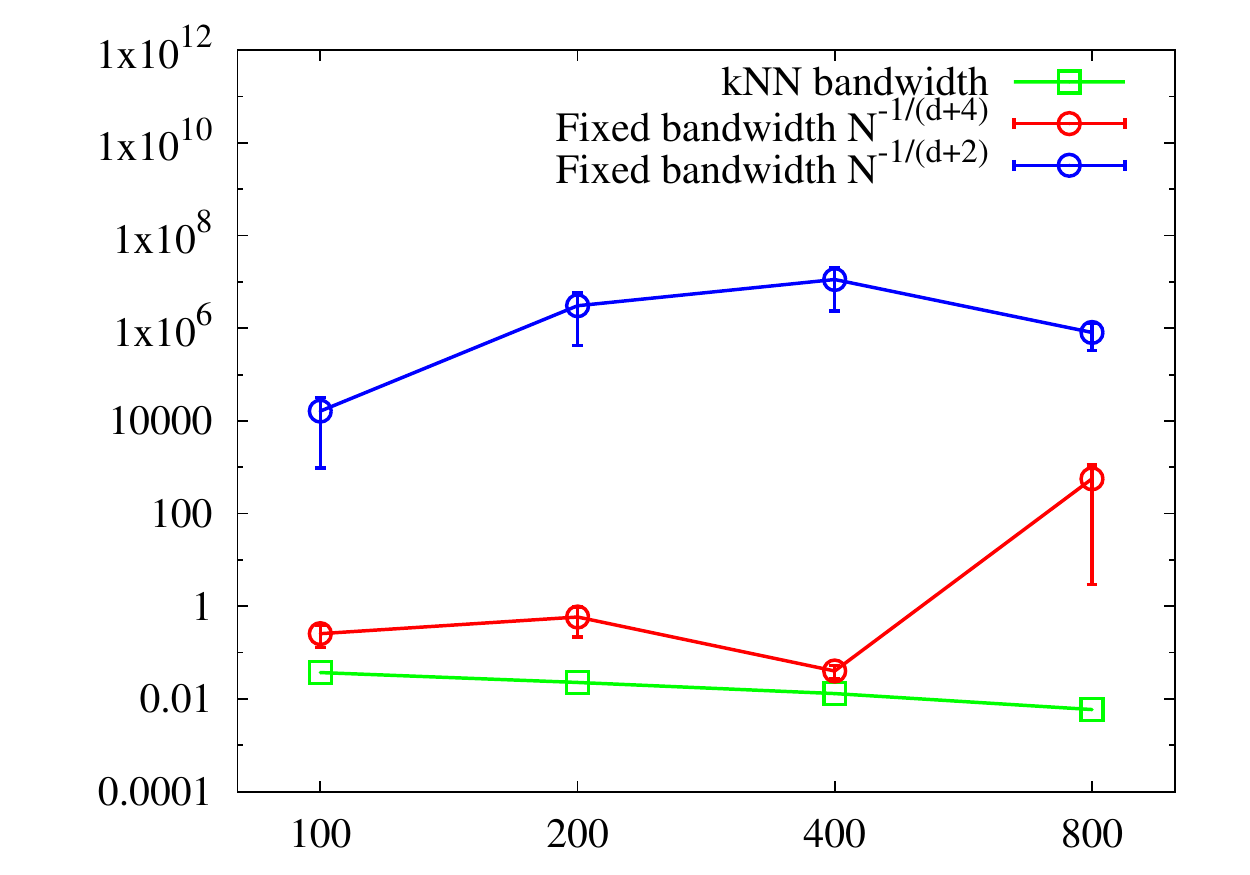}
	\put(-238,98){\small$ \E[(\hI-I)^2] $}
	\put(-139,-10){number of samples $n$}
	\end{center}
	\caption{Local and adaptive bandwidth  significantly improves over  rule-of-thumb fixed bandwidth.}
	\label{fig:bandwidth}
\end{figure}

The bottleneck in choosing such a small bandwidth is the bias,  scaling as $O(h^2 + (nh^d)^{-1}+ E_n)$ \cite{LWL12}, where the lower order dependence on $n$, dubbed $E_n$, is generally not known.
The barrier in choosing a {\em global} bandwidth of $h=O(n^{-1/d})$ is the strictly positive  bias whose value depends on the unknown distribution and cannot be subtracted off.
%, anda typical resolution is to resort back to a larger bandwidth $h=O(n^{-1/(d+2)})$ in the regime where $nh^d$ grows to infinity.
However, perhaps surprisingly,
%Instead, we enable the statistician to choose such a small bandwidth,
the proposed local and adaptive choice of the $k$-NN distance
admits an asymptotic bias that is independent of the unknown underlying distribution.
Manually subtracting off the
non-vanishing bias gives an asymptotically unbiased estimator, with a potentially faster convergence as numerically compared below.
Figure~\ref{fig:bandwidth} illustrates how $k$-NN based bandwidth significantly improves upon, say a rule-of-thumb choice of $O(n^{-1/(d+4)})$ explained above and another choice of $O(n^{-1/(d+2)})$.
In the left figure, we use the setting from Figure \ref{fig:entropy} (right) but with correlation $r=0.999$.
On the right, we generate $X\sim {\cal N}(0,1)$ and $U$ from uniform $[0,0.01]$ and let $Y=X+U$ and estimate $I(X;Y)$.
%We estimate mutual information $I(X;Y)$ from $n$ i.i.d.\ samples.
Following recent advances in \cite{LPS08,SP16}, % in extending $k$-NN entropy estimators to other functionals such as Renyi entropy ,
the proposed local estimator has a potential to be extended to, for example, Renyi entropy, but with a multiplicative bias  as opposed to additive.

% ------------------------------------------------------------------------------------------------------------------------

\section{Discussion}
\label{sec:order}
The topic of estimation of an integral functional of an unknown density from i.i.d.\ samples
  is a classical one in statistics and we tie together a few pertinent topics from the literature in the context of the results of this manuscript. 

\subsection{ Uniform order statistics and NN distances}
The expression for the asymptotic bias in \eqref{eq:defBias}
which is independent of the underlying distribution forms the main result of this paper and 
crucially depends on
Lemma \ref{lem:order_stat}.
Precisely, the lemma implies that the quantities $S_i$'s  in \eqref{eq:defS} converge in distribution to 
$\tS_i$'s in \eqref{eq:S}. There are two parts to this convergence result:
the nearest neighbor distances converge to uniform order statistics and
the directions to those nearest neighbors converge independently to Haar measures on the unit sphere.
The former has been extensively studied, for example see \cite{Rei12} for a survey of results.
The latter is a new result that we state in Lemma \ref{lem:order_stat}, and proved in Section \ref{sec:order_stat}. 
Intuitively, assuming smoothness,
the probability density $f_X$  in the  neighborhood of  a sample $X_i$ (as defined by the distance to the $k$-th nearest neighbor)
converges to a uniform distribution over a ball (of radius decreasing at the rate $\rho_{k,i}=\Theta(n^{-1/d})$), as more samples are collected.
The nearest neighbor distances and directions converge to those from the uniform distribution over the ball,
and  Lemma \ref{lem:order_stat} makes this intuition precise for the nearest $m$ neighbors up to  $m = O(n^{1/(2d)-\epsilon})$ with any arbitrarily small but positive $\varepsilon$.

Only the convergence analysis of the distances, and not the directions, is required for
traditional  $k$-NN based estimators, such as the
entropy estimator of \cite{KL87}. 
In the seminal paper,
\cite{KL87} introduced 
 {\em  resubstitution} entropy estimators of the form
$\hH(X) = -(1/n)\sum_{i=1}^n \log \hf_n(X_i)$ with
$\hf_n(x) =  {k}/(n\, C_d \,\rho_{k,x}^d)$ (as defined in \eqref{eq:knn}).
%We call this KL estimator for KOzachenko and Leonenko who first introduced it in \cite{KL87}.
 This $k$-NN estimator has a non-vanishing asymptotic bias, which was computed as
$B_{k,d}=(\psi(k)-\log(k))$ with the digamma function $\psi(\cdot)$ and was suggested to be manually removed.
 For $k=1$ this was proved in the original paper of \cite{KL87},
 which later was extended in \cite{SMH03,GLMN05} to general $k$.
This mysterious bias term $B_{k,d}=(\psi(k)-\log(k))$
whose original proofs in \cite{KL87,SMH03,GLMN05} provided little explanation for, can be alternatively proved
with both rigor and intuition
by making  connections to uniform order statistics.
For a special case of $k=1$, with extra assumptions on the support being compact,
such an elegant proof is provided in  \cite[Theorem 7.1]{BD16} which explicitly applies the
convergence of the nearest neighbor distance to uniform order statistics.
Namely,
\begin{eqnarray*}
	\E[\widehat{H}(X)] &=& \E\Big[\, - \frac1n \sum_{i=1}^n \log \Big(  \frac{k }{ n\, C_d \,\rho_{k,X_i}^d} \Big) \,\Big] \\
	&\to& \E\Big[ - \log \frac{k\,f(X_i)}{\sum_{j=1}^k E_j} \Big] \\
	&=& H(X) + \psi(k)-\log (k) \;,
\end{eqnarray*}
where
the asymptotic expression
follows from  $C_d \, n\, f(x) \rho_{k,x}^d \to \sum_{j=1}^k E_j$ as shown, for example, in Lemma \ref{lem:order_stat}
and we used $\E[\log \sum_{j=1}^k E_j]=\psi(k)$, where $\psi(k)=$ is the digamma function
defined as $\psi(x) = \Gamma^{-1}(x) d\Gamma(x)/dx$ and
for large $x$ it is approximately $\log(x)$ up to $O(1/x)$, i.e. $\psi(x)=\log x -1/(2x) + o(1/x)$.
Note that this only requires the convergence of the distance and not the direction.
Inspired by this modern approach, we extend such a connection in Lemma \ref{lem:order_stat}
to prove consistency of our estimator.

\subsection{ Convergence rate of the bias}
Establishing the convergence rate of the KL estimator is a challenging problem, and is not quite  resolved despite work 
 over the past three decades. 
The $O(1/n)$ convergence rate of the {\em variance}
 is established in \cite{BB83,LPS08,BD16,DF16} under various assumptions.
Establishing the convergence rate of the {\em bias} is more challenging.
It has been first studied in \cite{Hal84,Hal86}, where root-$n$ consistency
is shown in 1-dimension with bounded support and assuming  $f(x)$ is bounded below.
\cite{TV96} is the first to prove  a root mean squared error  convergence rate of $O(1/\sqrt{n})$
for general densities with  unbounded support in 1-dimension and exponentially decaying tail, such as the Gaussian density.
These assumptions are relaxed in \cite{EG09}, where zeroes and fat tails are allowed in $f(x)$.
In general $d$-dimensions,
 \cite{GOV16,SP16} prove bounds on the convergence rate of the bias for finite $k=O(1)$,
 and \cite{MML08,BSY16} for $k=\Omega(\log n)$.
Establishing the convergence rate for the bias of  the proposed local estimator is an interesting open problem --  it is interesting to see if the superior empirical performance of the local estimator is captured in the asymptotics of  rate of convergence of the bias. 

It is intuitive that kernel density estimators can capture the structure in the distribution
if the distribution lies on a lower dimensional manifold. 
This is made precise in \cite{OG09}, which also shows improved convergence rates 
for distributions whose support is on low dimensional manifolds. 
However, the estimator in \cite{OG09} critically uses  the 
geodesic distances between the sample points on the manifold.
Given that the proposed  estimators fit distributions locally,
a concrete question of interest is whether such an improvement can be achieved {\em without} 
such an explicit
knowledge of the geodesic distances, i.e., whether the local estimators automatically adapt to  underlying lower dimensional structures.

% Connections to manifolds and convergence rates there are also open questions.

%The  above $k$NN based entropy estimator has naturally been extended to a more general class of Renyi entropy estimators in  \cite{PPS10,LPS08,SP16}. The proposed local  entropy estimator can also be extended to estimate Renyi entropy, but with a multiplicative bias instead of additive.

\subsection{ Ensemble estimators} Recent works  \cite{SWH13,MH14,MSGH16,BSY16} have proposed ensemble estimators, 
which use known estimators based on kernel density estimators and $k$-NN methods and construct a new
estimate by taking the
weighted linear combination of those methods with varying bandwidth or $k$, respectively.
With a proper choice of the weights, which can be computed analytically by solving a simple linear program,
a boosting of the convergence rate can be achieved. The key property that allows the design of such ensemble estimators is that the leading terms (in terms of the sample size $n$) of the bias have a multiplicative constant that only depends on the unknown distribution. An intuitive explanation for this phenomenon is provided in \cite{BSY16} in the context of $k$-NN methods;  it is interesting to explore if such a phenomenon continues in the $k$-LNN scenario studied in this paper. Such a study would potentially lead to ensemble-based estimators in the local setting and also naturally allow a careful understanding of the rate of convergence of the bias term.

% ---------------------------------------------------------------------------------------------------------------------------------
\section{Proofs}

\subsection{Proof of proposition \ref{pro:closedform} }
\label{proof:prop}
We first prove the  derivation of the LLDE with degree $p=2$ in Equation \eqref{eq:p2}.
The gradient of the local likelihood evaluated at the  maximizer is zero \cite{Loa96}, which gives a computational tool for finding the maximizer:
\begin{align}
& \frac{1}{n} \sum_{j=1}^{n} K(\frac{X_j - x}{h}) = \int K(\frac{u-x}{h}) e^{a_0 + a_1^T (u-x) + (u-x)^T a_2 (u-x)} du \;,\label{eq:ll_0}\\
& \frac{1}{n} \sum_{j=1}^{n} \frac{X_j - x}{h} K(\frac{X_j - x}{h}) = \int \frac{u-x}{h} K(\frac{u-x}{h}) e^{a_0 + a_1^T (u-x) + (u-x)^T a_2 (u-x)} du \;,\label{eq:ll_1}\\
& \frac{1}{n} \sum_{j=1}^{n} \frac{(X_j - x)(X_j - x)^T}{h^2} K(\frac{X_j - x}{h}) \nonumber \\
&\;\;\;\; = \int \frac{(u-x)(u-x)^T}{h^2} K(\frac{u-x}{h}) e^{a_0 + a_1^T (u-x) + (u-x)^T a_2 (u-x)} du \;,\label{eq:ll_2}
\end{align}
where $K(x) = \exp\{-\|x\|^2/2\}$ is the Gaussian kernel.
Notice that the left-hand side of the equations are $S_0/n$, $S_1/n$ and $S_2/n$, respectively.
The RHS can be written in closed forms as:
\begin{eqnarray}
\frac{1}{n}S_0 &=& (2\pi)^{d/2} |{M}|^{-1/2} e^{a_0+\frac{1}{2}a_1^TM^{-1}a_1 } \;,\label{eq:ll'_0}\\
\frac{1}{n}S_1 &=& \frac{1}{nh} S_0  M^{-1} a_1\;,\label{eq:ll'_1}\\
\frac{1}{n}S_2 &=& \frac{1}{nh^2} S_0  (M^{-1} +  M^{-1}a_1a_1^TM^{-1})\;,\label{eq:ll'_2}
\end{eqnarray}
where $M = h^{-2} I_{d \times d} - 2a_2$ assuming $h$ sufficiently small such that $M$ is positive definite.
We want to derive $\hat{f}(x) = \exp\{a_0\}$ from the equations.
From \eqref{eq:ll'_1}  we get $M^{-1}a_1 = S_1(h/S_0)$.
Together with \eqref{eq:ll'_2},
we get $M^{-1} + M^{-1}a_1a_1^TM^{-1} = S_2(h^2/S_0)$.
Hence, $M^{-1} = (S_2/S_0 - (S_1/S_0) (S_1/S_0)^T) h^2 =  h^2 \Sigma $.
Plug them in~\eqref{eq:ll'_0}, we obtain the desired expression.
%\begin{eqnarray}
%\hat{f}(x) %&=& \exp\{a_0\} = \frac{S_0}{n (2\pi)^{d/2} (\det{M})^{-1/2}} \exp\{-\frac{1}{2} a_1^T M^{-1} a_1\} \,\notag\\
%&=&  \frac{S_0}{n (2\pi)^{d/2} (\det{M})^{-1/2}} \exp\{-\frac{1}{2} (M^{-1}a_1)^T M (M^{-1}a_1)\} \,\notag\\
%&=& \frac{S_0}{n (2\pi)^{d/2} |\Sigma|^{1/2} h^d} \exp\Big\{-\frac{1}{2S_0^2} (S_1^T \Sigma^{-1} S_1) \Big\} \;.
%\end{eqnarray}

% ---------------------------------------------------------------------------------------------------------------------------------
Analogously, for the derivation of the LLDE with degree $p=1$ in Equation \eqref{eq:p1}, we get
\begin{eqnarray}
\frac{1}{n}S_0 &=& (2\pi)^{d/2} h^{d} e^{a_0+\frac{h^2}{2}a_1^Ta_1 } \;,\label{eq:ll'_10}\\
\frac{1}{n}S_1 &=& \frac{h}{n} S_0   a_1\label{eq:ll'_11}\;.
\end{eqnarray}
This gives $a_1=(1/(hS_0))S_1$, and
$e^{a_0} = (S_0/(n(2\pi)^{d/2} h^{d}))\,\exp\{-0.5\|S_1\|^2/S_0^2\}$.

% -------------------------------------------------------------------------------------------------------------------------------------
\section{Proof of Lemma~\ref{lem:order_stat}}
\label{sec:order_stat}

Let us introduce some notations first.
Define $S^{d-1} \equiv \{x \in \mathbb{R}^d: \|x\| = 1\}$ as the unit $(d-1)$-dimensional sphere and $\sigma^{d-1}$ as a normalized spherical measure on $S^{d-1}$.
For any $\theta = (\theta_1, \dots, \theta_m) \in (S^{d-1})^m$ and $x = (x_1, \dots, x_m) \in \mathbb{R}_+^m$, define $\theta x \equiv (\theta_1 x_1, \dots, \theta_m x_m) \in \mathbb{R}^{d \times m}$.
For any set $B \in \mathbb{R}^{d \times m}$ and $\theta \in (S^{d-1})^m$, define $B_{\theta} = \{x \in \mathbb{R}_+^m: \theta x \in B\}$.
Let $\{\xi_i\}_{i=1}^m$ be i.i.d. random variables uniformly over $S^{d-1}$. Then for any joint random variables $(W_1, \dots, W_m) \in \mathbb{R}_+^m$ which are independent with $\{\xi_i\}_{i=1}^m$, we have
\begin{eqnarray}
\Pr\{ (\xi_1 W_1, \dots, \xi_m W_m) \in B\} &=& \int_{\theta \in (S^{d-1})^m } \Pr\{ (W_1, \dots, W_m) \in B_{\theta} \,|\, \theta\} \, d(\sigma^{d-1})^m(\theta) \;.
\end{eqnarray}

Let $Z=(Z_{1,i},\ldots,Z_{m,i})$, $\|Z\|=( \|Z_{1,i}\| ,\ldots, \|Z_{m,i}\|)$ and let $E=( E_1^{1/d} ,\ldots, (\sum_{\ell=1}^m E_\ell)^{1/d})$, then
\begin{eqnarray}
&&\left|\, \Pr\left\{\, (c_d nf(x))^{1/d} Z \in B \,\right\} - \Pr\left\{\,\left(\, \xi_1 E_1^{1/d}, \dots, \xi_m (\sum_{\ell=1}^m E_\ell)^{1/d} \,\right) \in B \,\right\} \,\right| \,\notag\\
&\leq&  \left|\, \Pr\left\{\, (c_d nf(x))^{1/d} Z \in B \,\right\} - \int_{\theta \in (S^{d-1})^m } \Pr\{ (E_1^{1/d}, \dots, (\sum_{\ell=1}^m E_\ell)^{1/d}) \in B_{\theta} \,|\, \theta\} \, d(\sigma^{d-1})^m(\theta) \,\right| \,\notag\\
&\leq& \left|\, \Pr\left\{\, (c_d nf(x))^{1/d} Z \in B \,\right\} - \int_{\theta \in (S^{d-1})^m } \Pr\{ (c_d nf(x))^{1/d} \|Z\| \in B_{\theta} \,|\, \theta\} \, d(\sigma^{d-1})^m(\theta) \,\right| \,\notag\\
&+& \int_{\theta \in (S^{d-1})^m} \left|\, \Pr\{ (c_d nf(x))^{1/d} \|Z\| \in B_{\theta} \,|\, \theta\} - \Pr\{ E \in B_{\theta} \,|\, \theta \} \,\right| d(\sigma^{d-1})^m(\theta) \;.\label{eq:tv}
\end{eqnarray}

Now consider the first term in~\eqref{eq:tv}. %To simplify the notations, let $\mathcal{E}_{\lambda}$ be the event of $\left(\, 2nf(x)Z_{1,i}, \dots, 2nf(x)Z_{m,i} \,\right) \in B_{\lambda}$ and $\mathcal{E}'_{\lambda}$ be the event of $\left(\, 2nf(x)|Z_{1,i}|, \dots, 2nf(x)|Z_{m,i}| \,\right) \in |B_{\lambda}|$.
We consider two cases separately.

{\bf Case 1.} If $\|Z_{m,i}\| \geq (\sqrt{n} c_d f(x))^{-1/d}$, we show that the tail events happen with a low probability. Denote $B(x,r) = \{z: \|z-x\| \leq r\}$ and let $p = \Pr\{t \in B(x, \|Z_{m,i}\|)\} = \int_{B(x,\|Z_{m,i}\|)} f(t) dt$. Since $f$ is twice continuously differentiable, we can see that $p \geq 0.5 c_d \|Z_{m,i}\|^d f(x) \geq 0.5/\sqrt{n}$ for sufficiently large $n$. Therefore,
\begin{eqnarray}
\Pr\{\|Z_{m,i}\| \geq (\sqrt{n} c_d f(x))^{-1/d}\} &=& \sum_{\ell=0}^{m-1} {n \choose \ell} p^{\ell} (1-p)^{n-\ell} \leq \sum_{\ell=0}^{m-1} n^{\ell} \Big(1-\frac{1}{2\sqrt{n}}\Big)^{(n-\ell)} \,\notag\\
&\leq& \sum_{\ell=0}^{m-1} n^l e^{-(\sqrt{n} - \ell\sqrt{n})/2} \leq m n^m e^{-(\sqrt{n} - m/\sqrt{n})/2} \;.
\end{eqnarray}

{\bf Case 2.} If $\|Z_{m,i}\| < (\sqrt{n} c_d f(x))^{-1/d}$, let $\overline{B} = \{t: (c_d n f(x))^{1/d} t \in B \textrm{ and } \|t_m\| < (\sqrt{n} c_d f(x))^{-1/d} \}$ and $\overline{B_{\theta}} = \{t: (c_d n f(x))^{1/d} t \in B_{\theta} \textrm{ and } t_m < (\sqrt{n} c_d f(x))^{-1/d} \}$.
Note that 
\begin{eqnarray}
\Pr (Z \in \tA) = (n!/(n-k)!)\int_{t\in\tA} \prod_{j=1}^m f(x+t_j) \Pr_X(|X-x|>|t_m|)^{n-m} dt \;,
\end{eqnarray}
which gives
\begin{eqnarray}
&& \frac{\int_{\theta \in (S^{d-1})^m} \Pr\{(c_d n f(x))^{1/d}\|Z\| \in B_{\theta}, \|Z_{m,i}\| < (\sqrt{n} c_d f(x))^{-1/d} \,|\,\theta\} \, d(\sigma^{d-1})^m(\theta)}{\Pr\{(c_d n f(x))^{1/d} Z \in B, \|Z_{m,i}\| < (\sqrt{n} c_d f(x))^{-1/d}\}} \,\notag\\
&=& \frac{\int_{\theta \in (S^{d-1})^m} \Pr\{\|Z\| \in \overline{B_{\theta}} \,|\,\theta\} \, d(\sigma^{d-1})^m(\theta)}{\Pr\{ Z \in \overline{B}\}} \,\notag\\
&=& \frac{\int_{\theta \in (S^{d-1})^m} \frac{n!}{(n-k)!} \left(\, \int_{t \in \overline{B_{\theta}}} \left(\, \prod_{j=1}^m f(x + \theta_j t_j) \,\right)\, \left(\, \Pr\{\|X-x\| > \|t_m\|\} \,\right)^{n-m} dt \,\right)\, d(\sigma^{d-1})^m(\theta) }{ \frac{n!}{(n-k)!} \int_{t \in \overline{B}} \left(\, \prod_{j=1}^m f(x + t_j) \,\right)\, \left(\, \Pr\{\|X-x\| > \|t_m\|\} \,\right)^{n-m} dt} \,\notag\\
&\leq& \frac{\sup_{\theta \in (S^{d-1})^m} \sup_{t \in \overline{B_{\theta}}} \prod_{j=1}^m f(x+ \theta_j t_j)}{\inf_{t \in \overline{B}} \prod_{j=1}^m f(x+t_j)} \, \notag\\
&\leq& \left(\, \frac{\sup_{\|t\| \leq (\sqrt{n} c_d f(x))^{-1/d}} f(x+t)}{\inf_{\|t\| \leq (\sqrt{n} c_d f(x))^{-1/d}} f(x+t )}\,\right)^m \;,\label{eq:ratio}
\end{eqnarray}

where the first inequality follows from the fact that
$\int_{\theta \in (S^{d-1})^m} ( \int_{\overline{B_\theta}} g(t_m)dt ) d (\sigma^{d-1})^m (\theta) = \int_{\overline{B}} g(\|t_m\|)dt$.
Since $f$ is continuously differentiable, by mean value theorem, there exists $a, b \in B(x, (\sqrt{n} c_d f(x))^{-1/d})$ such that
\begin{eqnarray}
\frac{\sup_{\|t\| \leq (\sqrt{n} c_d f(x))^{-1/d}} f(x+t)}{\inf_{\|t\| \leq (\sqrt{n} c_d f(x))^{-1/d}} f(x+t )} &=& \frac{f(b) + (a-b)^T \nabla f(a) }{f(b)} \leq 1 + \frac{2 (\sqrt{n} c_d f(x))^{-1/d} \|\nabla f(a)\|}{f(b)} \;,
\end{eqnarray}
By the assumption, there exists a ball $B(x,\varepsilon)$ such that $\|\nabla f(a)\| = O(1)$ and $f(a) > 0$ for all $a \in B(x, \varepsilon)$, so for sufficiently large $n$ such that $(\sqrt{n} c_d f(x))^{-1/d} < \varepsilon$, there exists some constant $C$ such that $\sup_{\|t\| \leq (\sqrt{n} c_d f(x))^{-1/d}} f(x+t) \leq (1 + C n^{-1/(2d)}) \inf_{\|t\| \leq (\sqrt{n} c_d f(x))^{-1/d}} f(x+t )$. Therefore,~\eqref{eq:ratio} is upper bounded by $(1 + C n^{-1/(2d)})^m$. Similarly,~\eqref{eq:ratio} is lower bounded by $(1 - C n^{-1/(2d)})^m$.

For simplicity, let $\mathcal{E} = \{\|Z_{m,i}\| < (\sqrt{n} c_d f(x))^{-1/d}\}$. Then combining the two cases, the first term in~\eqref{eq:tv} is bounded by:
\begin{eqnarray}
&& \left|\, \Pr\left\{\, (c_d nf(x))^{1/d} Z \in B \,\right\} - \int_{\theta \in (S^{d-1})^m } \Pr\{ (c_d nf(x))^{1/d} \|Z\| \in B_{\theta} \,|\, \theta\} \, d(\sigma^{d-1})^m(\theta) \,\right| \,\notag\\
&\leq& \Pr\left\{\, (c_d nf(x))^{1/d} Z \in B, \mathcal{E}^C \,\right\} + \int_{\theta \in (S^{d-1})^m } \Pr\{ (c_d nf(x))^{1/d} \|Z\| \in B_{\theta}, \mathcal{E}^C \,|\, \theta\} \, d(\sigma^{d-1})^m(\theta) \,\notag\\
&+& \left|\, \Pr\left\{\, (c_d nf(x))^{1/d} Z \in B, \mathcal{E}\,\right\} - \int_{\theta \in (S^{d-1})^m } \Pr\{ (c_d nf(x))^{1/d} \|Z\| \in B_{\theta}, \mathcal{E} \,|\, \theta\} \, d(\sigma^{d-1})^m(\theta) \,\right| \,\notag\\
&\leq& \Pr\{\mathcal{E}^C\} + \int_{\theta \in (S^{d-1})^m } \Pr\{ \mathcal{E}^C \} \, d(\sigma^{d-1})^m(\theta) \,\notag\\
&+& \Pr\left\{\, (c_d nf(x))^{1/d} Z \in B, \mathcal{E}\,\right\} \, \left|\, 1 -  \frac{\int_{\theta \in (S^{d-1})^m } \Pr\{ (c_d nf(x))^{1/d} \|Z\| \in B_{\theta}, \mathcal{E} \,|\, \theta\} \, d(\sigma^{d-1})^m(\theta)}{\Pr\left\{\, (c_d nf(x))^{1/d} Z \in B, \mathcal{E}\,\right\}}\,\right| \,\notag\\
&\leq& 2 \Pr\{\mathcal{E}^C\} + \Pr\left\{\, (c_d nf(x))^{1/d} Z \in B, \mathcal{E}\,\right\} \max\{(1+Cn^{-1/(2d)})^m - 1, 1 - (1-Cn^{-1/(2d)})^m\} \,\notag\\
&\leq& 2 m n^m e^{-(\sqrt{n} - m/\sqrt{n})/2} + \max\{(1+Cn^{-1/(2d)})^m - 1, 1 - (1-Cn^{-1/(2d)})^m\} \;.\label{eq:ub100}
\end{eqnarray}

Now consider the second term of~\eqref{eq:tv}. We will use Corollary 5.5.5 of ~\cite{Rei12} to show that
this term vanishes for $m=O(\log n)$ and as $n$ grows.

\begin{lemma}[Corollary 5.5.5, ~\cite{Rei12}]
    \label{lem:reiss}
    Let $Y_1, Y_2, \dots, Y_n$ be i.i.d. samples from unknown distribution with pdf $f$. Let $Y_{1:n} \leq Y_{2:n} \leq \dots \leq Y_{n:n}$ be the order statistics. Assume the density $f$ satisfies $|\log f(y)| \leq L y^{\delta}$ for $0 < y < y_0$ and $f(y) = 0$ for $y < 0$, where $L$ and $\delta$ are constants. Then
    \begin{eqnarray}
        d_{\rm TV}\Big(\, n\,\left(\,Y_{1:n}, Y_{2:n}, \dots, Y_{m:n}\right) , \big(\,E_1, E_1+E_2, \dots, \sum_{j=1}^m E_j\big) \,\Big)  \leq C_0 \left(\, (m/n)^{\delta} m^{1/2} + m/n \,\right) \;,
    \end{eqnarray}
    where $C_0 > 0$ is a constant. $E_1, \dots, E_m$ are i.i.d standard exponential random variables.
\end{lemma}

Now for fixed $x$, consider the distribution of $c_d f(x)\|X - x\|^d$ denoted by $\tilde{P}$. Define $Y_1, Y_2, \dots, Y_n$ drawn i.i.d. from $\tilde{P}$. We can see that $c_d f(x) \|Z\|^d \stackrel{\mathcal{L}}{=} (Y_{1:n}, \dots, Y_{m:n})$, where $\stackrel{\mathcal{L}}{=}$ denotes equivalence in distribution. The pdf $\tilde{f}$ of $\tilde{P}$ is given by:
\begin{eqnarray}
\tilde{f}(t) = \frac{d}{dt} \Pr\{c_d f(x)\|X - x\|^d \leq t\} = \frac{d}{dt} \int_{y \in B(x, r_t)} f(y) dy  \;.
\end{eqnarray}
where $r_t = (t/(c_d f(x)))^{1/d}$. Here we have:
\begin{eqnarray}
\frac{dr_t}{dt} = \frac{t^{1/d-1} (c_d f(x))^{-1/d}}{d} = \frac{1}{f(x) d c_d r_t^{d-1}} \;.
 \end{eqnarray}
 If $f$ is twice continuously differentiable, we have:
\begin{eqnarray}
\left|\, \tilde{f}(t) - 1 \,\right| &=& \left|\, \frac{d}{dt} \int_{y \in B(x, r_t)} f(y) dy - 1 \,\right| = \left|\, \frac{dr_t}{dt} (\frac{d}{dr_t} \int_{y \in B(x, r_t)} f(y) dy) - 1 \,\right| \,\notag\\
&=& \frac{1}{f(x) d c_d r_t^{d-1}} \left|\, \frac{d}{dr_t} \left(\, \int_{y \in B(x,r_t)} f(y) dy \,\right)- f(x) d c_d r_t^{d-1} \,\right| \,\notag\\
&=& \frac{1}{f(x) d c_d r_t^{d-1}} \left|\, \int_{y \in S^{d-1}(x,r_t)} (f(y) - f(x)) d\sigma^{d-1}(y) \,\right| \;,
\end{eqnarray}
where $S^{d-1}$ is the $(d-1)$-sphere centered at $x$ with radius $r_t$ and $\sigma^{d-1}$ is the spherical measure. By mean value theorem, there exists $a(y) \in B(x,r_t)$ such that $f(y) - f(x) = (y-x)^T \nabla f(x) + (a(y)-x)^T H_f(a(y)) (a(y)-x)$, where $a(y)$ depends on $y$. Therefore,
\begin{eqnarray}
&& \left|\, \int_{y \in S^{d-1}(x,r_t)} (f(y) - f(x)) d\sigma^{d-1}(y) \,\right| \,\notag\\
&=& \left|\, \underbrace{\int_{y \in S^{d-1}(x,r_t)} (y-x)^T \nabla f(x)  d\sigma^{d-1}(y)}_{=0} + \int_{y \in S^{d-1}(x,r_t)} (a(y)-x)^T H_f(a(y)) (a(y)-x)  d\sigma^{d-1}(y)\,\right| \,\notag\\
&\leq& \left(\, \sup_{a \in B(x,r_t)} \|H_f(a)\| \, \|a-x\|^2 \,\right) \sigma^{d-1}(S^{d-1}(x,r_t)) \,\notag\\
&\leq& d c_d r_t^{d+1} \left(\, \sup_{a \in B(x,r_t)} \|H_f(a)\| \,\right)
\end{eqnarray}
Since there exists a ball $B(x,\varepsilon)$ such that $\|H_f(a)\| = O(1)$ for all $a \in B(x, \varepsilon)$. Therefore, for sufficiently small $t$ such that $r_t < \varepsilon$, we have:
\begin{eqnarray}
\left|\, \tilde{f}(t) - 1 \,\right| \leq \frac{d c_d r_t^{d+1} \left(\, \sup_{a \in B(x,r_t)} \|H_f(a)\| \,\right)}{f(x) d c_d r_t^{d-1}} = \frac{r_t^2 \left(\, \sup_{a \in B(x,r_t)} \|H_f(a)\| \,\right)}{f(x)} \;.
\end{eqnarray}
Recall that $r_t = (t/(c_df(x)))^{1/d}$, so there exists $L > 0$ such that $|\tilde{f}(t) - 1| \leq L t^{2/d}$ for sufficiently small $t$. Hence, $|\log \tilde{f}(t)| \leq L' t^{2/d}$ for some $L'>0$ and sufficiently small $t$.
 So $\tilde{f}$ satisfies the condition in Lemma.~\ref{lem:reiss} with $\delta = 2/d$. Therefore, for any $B_{\theta} \subseteq \mathbb{R}_+^m$, we have:
\begin{eqnarray}
&&\left|\, \Pr\{ (c_d nf(x))^{1/d} \|Z\| \in B_{\theta} \} - \Pr\{ E \in B_{\theta}\} \,\right| \,\notag\\
&\leq& d_{\rm TV} \left(\, c_d n f(x) \|Z\|^d, \big(\,E_1, E_1+E_2, \dots, \sum_{j=1}^m E_j\big) \,\right) \,\notag\\
&\leq& C_0\left(\, (\frac{m}{n})^{2/d} m^{1/2} + \frac{m}{n} \,\right) \;.\label{eq:ub200}
\end{eqnarray}
Therefore, by combing~\eqref{eq:ub100} and~\eqref{eq:ub200}, we have:
\begin{eqnarray}
&&\left|\, \Pr\left\{\, (c_d nf(x))^{1/d} Z \in B \,\right\} - \Pr\left\{\,\left(\, \xi_1 E_1^{1/d}, \dots, \xi_m (\sum_{l=1}^m E_{\ell})^{1/d} \,\right) \in B \,\right\} \,\right| \,\notag\\
&\leq& 2 m n^m e^{-\frac{\sqrt{n} - m/\sqrt{n}}{2}} + \max\{(1+Cn^{\frac{-1}{2d}})^m - 1, 1 - (1-Cn^{\frac{-1}{2d}})^m\} + C_0\left(\, (\frac{m}{n})^{\frac{2}{d}} m^{\frac{1}{2}} + \frac{m}{n} \,\right)\;,
\end{eqnarray}
for any set $B \in \mathbb{R}^{d \times m}$. Therefore, the total variation distance $d_{\rm TV} ( (c_d nf(x))^{1/d} (\, Z_{1,i}, Z_{2,i}, \dots, Z_{m,i} ) ,( \xi_1 E_1^{1/d}, \xi_2(E_1 + E_2)^{1/d}, \dots , \xi_{m}(\sum_{\ell=1}^{m} E_\ell)^{1/d} \,) )$ is bounded by the RHS quantity. By taking $m = O(\log n)$, the RHS converges to 0 as $n$ goes to infinity. Therefore, we have the desired statement.

% ---------------------------------------------------------------------------------------------------------------------------------
\section{Proof of Theorem \ref{thm:unbiased}}
\label{sec:proofmaintheorem}
We first compute the asymptotic bias.
We define new notations to represent the estimate as
\begin{eqnarray*}
	\hH^{(n)}_k=\frac1n \sum_{i=1}^n \Big\{ \underbrace{h\big(\,(c_d nf(X_i))^{1/d} Z_{k,i}, S_{0,i}, S_{1,i}, S_{2,i})\,\big) - \log f(X_i)}_{\equiv H_i}  \Big\} \;,
\end{eqnarray*}
where  $h : \mathbb{R}^d \times \mathbb{R} \times \mathbb{R}^d \times \mathbb{R}^{d \times d} \to \mathbb{R} $ is defined as 
\begin{align}
&h(t_1, t_2, t_3, t_4) = \nonumber\\
&\;\;d \log \|t_1\| + d \log (2 \pi) - \log c_d - \log t_2  + \frac{1}{2} \log \left(\, \det\left(\, \frac{t_4}{t_2} - \frac{t_3 t_3^T}{t^2_2}\,\right) \,\right)+ \frac{1}{2} t_3^T (t_4 - t_3 t_3^T)^{-1} t_3 \;.
\end{align}
%and $\bar{h} = h$ if $|h| < 10^{10}$, $\bar{h} = 10^{10}$ if $h > 10^{10}$ and $\bar{h} = -10^{10}$ if $h < -10^{10}$. Such a truncation is necessary for exchanging the limit.
%Therefore,  $\bar{h}( 2nf(x)\rho_{k,i}, S_{0,i}, S_{1,i}, S_{2,i}) = H_i + \log f(x)$.
Let $H_i \equiv h((c_dnf(X_i))^{1/d}Z_{k,i}, S_{0,i}, S_{1,i}, S_{2,i}))  - \log f(X_i)$.
 Since the terms $H_1, H_2, \dots, H_n$ are identically distributed, the expected value of $\hH_k^{(n)}$ converges to
\begin{eqnarray}
	 \label{eq:entropy1}
    \lim_{n \to \infty} \E[\hH_k^{(n)}] &=& \lim_{n \to \infty} \E[H_1]  \;\;= \;\; \lim_{n\to \infty} \E_{X_1}\big[\E[H_1|X_1]\big]\;\;\,  
\end{eqnarray}
Typical approach of dominated convergence theorem cannot be applied to the above limit, 
since analyzing $\E[H_1|X_1]$ for finite sample $n$ is challenging. 
In order to exchange the  limit with the (conditional) expectation, we assume the following Ansatz \ref{ansatz} to be true.  
As noted in \cite{PPS10} this is  common  in the literature on consistency of $k$-NN estimators, 
where the same assumptions have been implicitly made without explicitly stating as such, 
in existing analyses of entropy estimators including \cite{KL87,GLMN05,LPS08,WKV09}. 
This assumption can be avoided for Renyi entropy as in the proof of consistency in \cite{PPS10} or 
for sharper results such as the convergence rate of the bias with respect to the sample size but with more assumptions as in \cite{GOV16,SP16,BSY16}. 
\begin{ansatz}
	The following exchange of limit holds: 
	\label{ansatz}
	\begin{eqnarray}
    	\lim_{n \to \infty} \E[H_1] \;\; =\;\; \E_{X_1} \left[\, \lim_{n \to \infty} \E[H_1|X_1] \,\right] \,,
	\end{eqnarray}
\end{ansatz}
%where we apply the dominated convergence theorem to exchange the limit and expectation over $X_1$, since  $|H_1| < 10^{10}$.
Under this ansatz, perhaps surprisingly, we will show that the expectation inside converges to $-\log f(X_1)$ plus some bias that is independent of the underlying distribution. Precisely, for almost every $x$ and given $X_1 = x$,
\begin{eqnarray}
\E[H_1 | X_1 = x] + \log f(x) &=& \E \left[\, h ( (c_d nf(x))^{1/d} Z_{k,i}, S_{0,1}, S_{1,i}, S_{2,i}) \,\right] \,\notag\\
& \longrightarrow & B_{k,d}\;,
\label{eq:entropy2}
%\E\left[\, \bar{h} \left(\,\xi_k \sum_{l=1}^k E_\ell, \tilde{S}^{(\infty)}_{0,i}, \tilde{S}^{(\infty)}_{1,i}, \tilde{S}^{(\infty)}_{2,i} \,\right) \,\right] \;,
\end{eqnarray}
as $n\to \infty$ where $B_{k,d}$  is a constant  that only depends on $k$ and $d$, defined in \eqref{eq:defB}. This implies that
\begin{eqnarray}
    \E_{X_1} \left[\, \lim_{n \to \infty} \E[H_1|X_1]\right] & =&     \E_{X_1} [-\log f(X_1) + B_{k,d}] \,\notag\\
    &=& H(X) + B_{k,d} \;.\label{eq:entropy4}
\end{eqnarray}
Together with \eqref{eq:entropy1}, this finishes the proof of the desired claim.

We are now left to prove the convergence of  \eqref{eq:entropy2}. We first give a formal definition of
the bias $B_{k,d}$ by replacing the sample defined quantities by a similar quantities defined from order-statistics,
and use Lemma \ref{lem:order_stat} to prove the convergence.
Recall that our order-statistics is defined by two sequences of $m$ i.i.d. random variables:
i.i.d. standard exponential random variables $E_1, \dots, E_m$ and
i.i.d. random variables  $\xi_1, \dots, \xi_m$ uniformly distributed over $S^{d-1}$.
We define
\begin{eqnarray}
	\label{eq:defB}
	B_{k,d} &\equiv &
	\E\left[\, {h} \left(\,\xi_k \left(\, \sum_{\ell=1}^k E_\ell \,\right)^{1/d}, \tilde{S}^{(\infty)}_{0}, \tilde{S}^{(\infty)}_{1}, \tilde{S}^{(\infty)}_{2} \,\right) \,\right] \;,
\end{eqnarray}
where, as we will show,
$\tS_\alpha^{(\infty)}$ is the limit of empirical quantity $S_{\alpha,i}$ defined from samples for each $\alpha\in\{0,1,2\}$,
and we know that $(c_d nf(x))^{1/d} Z_{k,i}$ converges to $\xi_k (\sum_{\ell=1}^k E_\ell)^{1/d}$ for almost every $x$ from Lemma \ref{lem:order_stat}.
$S^{(\infty)}$ is defined by a convergent random sequence
\begin{eqnarray}
\tilde{S}^{(m)}_{\alpha}  &\equiv&  \sum_{j=1}^{m} \frac{\xi_j^{(\alpha)} (\sum_{\ell=1}^j E_\ell)^{\alpha/d} }{(\sum_{\ell=1}^k E_\ell)^{\alpha/d} } \exp\Big\{- \frac{(\,\sum_{\ell=1}^j E_\ell\,)^{2/d}}{2(\,\sum_{\ell=1}^k E_\ell \,)^{2/d}} \Big\} \;,
\end{eqnarray}
where $\xi_j^{(0)} = 1$, $\xi_j^{(1)} = \xi_j$, $\xi_j^{(2)} = \xi_j \xi_j^T$ and $\tS^{(\infty)}_\alpha = \lim_{m\to \infty} \tS^{(m)}_\alpha$.
This limit exists, since $\tS_0^{(m)}$ is non-decreasing in $m$,
and the convergence of $\tS_1^{(m)}$  and $\tS_2^{(m)}$ follows from Lemma  \ref{lem:tail}.
We introduce simpler notations for the joint random variables:
$\tS^{(m)}=(\xi_k (\sum_{\ell=1}^k E_{\ell})^{1/d}, \tilde{S}^{(m)}_{0},\tilde{S}^{(m)}_{1},\tilde{S}^{(m)}_{2})$ and
 $\tS^{(\infty)}=(\xi_k (\sum_{\ell=1}^k E_{\ell})^{1/d}, \tilde{S}^{(\infty)}_{0}, \tilde{S}^{(\infty)}_{1}, \tilde{S}^{(\infty)}_{2})$.
Considering the quantities  $S^{(n)} = ((c_d nf(x))^{1/d} Z_{k,i}, S_{0,i}, S_{1,i}, S_{2,i})$
defined from samples, we show that this converges
to $\tS^{(\infty)}$.
Precisely, applying  triangular inequality,
\begin{eqnarray}
	d_{\rm TV} (S^{(n)},\tS^{(\infty)}) &\leq&
	d_{\rm TV} (S^{(n)},\tS^{(m)}) + d_{\rm TV} (\tS^{(m)},\tS^{(\infty)}) \;, \label{eq:entropy3}
%	&=&d_{\rm TV} (S^{(n)},\tS^{(m)})  + d_{\rm TV}((\tS_0^{(m)},\tS_1^{(m)},\tS_2^{(m)}),(\tS_0^{(\infty)},\tS_1^{(\infty)},\tS_2^{(\infty)})) \nonumber\;.,
\end{eqnarray}
%&& \Big\| (2nf(x) \rho_{k,i}, S_{0,i}, S_{1,i}, S_{2,i}) - (\xi_k \sum_{l=1}^k E_\ell, \tilde{S}^{(\infty)}_{0,i}, \tilde{S}^{(\infty)}_{1,i}, \tilde{S}^{(\infty)}_{2,i})\Big\|_{TV} \,\notag\\
%&\leq& \Big\| (2nf(x) \rho_{k,i}, S_{0,i}, S_{1,i}, S_{2,i}) - (\xi_k \sum_{l=1}^k E_\ell, \tilde{S}^{(m)}_{0,i}, \tilde{S}^{(m)}_{1,i}, \tilde{S}^{(m)}_{2,i})\Big\|_{TV} \,\notag\\
%&+& \|\,(\xi_k \sum_{l=1}^k E_\ell, \tilde{S}^{(m)}_{0,i},\tilde{S}^{(m)}_{1,i},\tilde{S}^{(m)}_{2,i}) - (\xi_k \sum_{l=1}^k E_\ell, \tilde{S}^{(\infty)}_{0,i},\tilde{S}^{(\infty)}_{1,i},\tilde{S}^{(\infty)}_{2,i})\,\|_{TV} \,\notag\\
%&=& \Big\| (2nf(x) \rho_{k,i}, S_{0,i}, S_{1,i}, S_{2,i}) - (\xi_k \sum_{l=1}^k E_\ell, \tilde{S}^{(m)}_{0,i}, \tilde{S}^{(m)}_{1,i}, \tilde{S}^{(m)}_{2,i})\Big\|_{TV} \,\notag\\
%&+& \|\,( \tilde{S}^{(m)}_{0,i},\tilde{S}^{(m)}_{1,i},\tilde{S}^{(m)}_{2,i}) - ( \tilde{S}^{(\infty)}_{0,i},\tilde{S}^{(\infty)}_{1,i},\tilde{S}^{(\infty)}_{2,i})\,\|_{TV} \,\notag\\
%&\stackrel{n \rightarrow \infty}{\longrightarrow}& 0 \label{eq:converge}\;,
%\end{eqnarray}
and we show that both terms converge to zero for any $m = \Theta(\log n)$.
Given that $\bh$ is continuous and bounded, this implies that
\begin{eqnarray*}
	\lim_{n\to\infty} \E[H_1|X_1 = x] &=& \E[\lim_{n\to\infty} \bh(S^{(n)})-\log f(x)|X_1 = x] \\
	&=& -\log f(x) + \E[\bh(\tS^{(\infty)})]\;,
\end{eqnarray*}
for almost every $x$, proving \eqref{eq:entropy4}.

%and $\bar{h}$ is a fixed function. Therefore, RHS is only depend on $k$. Let it be $C_{k}$, then we know that $\lim_{n \to \infty} \E[H_i | X_i = x] = - \log f(x) + C_{k}$.

The convergence of the first term follows from Lemma \ref{lem:order_stat}.
Precisely, consider the function $g_{m}: \mathbb{R}^{d \times m} \to \mathbb{R}^d \times \mathbb{R} \times \mathbb{R}^d \times \mathbb{R}^{d \times d}$ defined as:
\begin{eqnarray}
g_{m}(t_1, t_2, \dots, t_{m}) = \left(\, t_k, \sum_{j=1}^{m} \exp\{-\frac{\|t_j\|^2}{2\|t_k\|^2}\}, \sum_{j=1}^{m} \frac{t_j}{\|t_k\|} \exp\{-\frac{\|t_j\|^2}{2\|t_k\|^2}\}, \sum_{j=1}^{m} \frac{t_j t_j^T}{\|t_k\|^2} \exp\{-\frac{\|t_j\|^2}{2\|t_k\|^2}\} \,\right) \;,
\end{eqnarray}
such that $S^{(n)} = g_{m} \left(\, (c_dnf(x))^{1/d} \left(\, Z_{1,i}, Z_{2,i}, \dots, Z_{m,i} \,\right) \,\right)$
, which follows from the definition of
$S^{(n)}=((c_dnf(x))^{1/d}Z_{k,i},S_{0,i},S_{1,i},S_{2,i})$ in \eqref{eq:defS}.
Similarly,
%For $\alpha \in \{0, 1, 2\}$, define random variables
%\begin{eqnarray}
%\tilde{S}^{(m)}_{\alpha,i} = \sum_{j=1}^{m} \frac{(\xi_j\sum_{l=1}^j E_{\ell})^{\alpha}}{|\,\xi_k \sum_{l=1}^k E_\ell\,|^{\alpha}} \exp\{- \frac{(\,\sum_{l=1}^j E_{\ell}\,)^{2}}{2(\,\sum_{l=1}^k E_\ell\,)^{2}}\} \label{eq:S}\;,\\
%\tilde{T}^{(m)}_{\alpha,i} = \sum_{j=m+1}^{\infty} \frac{(\xi_j\sum_{l=1}^j E_{\ell})^{\alpha}}{|\,\xi_k \sum_{l=1}^k E_\ell\,|^{\alpha}} \exp\{- \frac{(\,\sum_{l=1}^j E_{\ell}\,)^{2}}{2(\,\sum_{l=1}^k E_\ell\,)^{2}}\} \label{eq:T}\;,
%\end{eqnarray}
%such that
 $\tS^{(m)} = g_{m} \left(\, \xi_1 E_1^{1/d}, \xi_2(E_1 + E_2)^{1/d}, \dots \xi_{m}(\sum_{\ell=1}^{m} E_\ell)^{1/d} \,\right)$.
 %$\tS^{(\infty)}\tilde{S}^{(m)}_{\alpha,i} + \tilde{T}^{(m)}_{\alpha,i} = \tilde{S}^{(\infty)}_{\alpha,i}$.
 Since $g_{m}$ is continuous, so for any set $A \in \mathbb{R}^d \times \mathbb{R} \times \mathbb{R}^d \times \mathbb{R}^{d \times d}$, there exists a set $\tA \in \mathbb{R}^{d \times m}$ such that $g_{m}(\tA) = A$. So for any $x$ such that there exists $\varepsilon > 0$ such that $f(a) > 0$, $\|\nabla f(a)\| = O(1)$ and $\|H_f(a)\| = O(1)$ for any $\|a - x\| < \varepsilon$, we have:
\begin{align}
& d_{\rm TV} ( S^{(n)},\tS^{(m)})  \,\notag\\
&= \sup_{A } %\sup_{A \in \mathbb{R}^d \times \mathbb{R} \times \mathbb{R}^d \times \mathbb{R}^{d \times d}}
\left|\, \Pr\left\{g_{m} \left(\, (c_dnf(x))^{\frac1d} Z_{1,i},  \dots, (c_dnf(x))^{\frac1d}Z_{m,i} \,\right) \in A \right\} -
	\Pr\{g_{m} (\, \xi_1 E_1^{\frac1d}, \dots \xi_{m}(\sum_{l=1}^{m} E_{\ell})^{\frac1d} \,) \in A \} \,\right| \,\notag\\
&\leq \sup_{\tA \in \mathbb{R}^{d \times m}} \left|\, \Pr\left\{\left(\, (c_dnf(x))^{1/d} Z_{1,i},  \dots, (c_dnf(x))^{1/d} Z_{m,i} \,\right) \in \tA \right\} - \Pr\{(\, \xi_1 E_1^{1/d},  \dots \xi_{m} (\sum_{\ell=1}^{m} E_\ell)^{1/d} \,) \in \tA \} \,\right| \,\notag\\
&= d_{\rm TV}\left( \left(\, (c_dnf(x))^{1/d} Z_{1,i},  \dots, (c_dnf(x))^{1/d} Z_{m,i} \,\right) \,,\, \left(\, \xi_1 E_1^{1/d}, \dots \xi_{m} (\sum_{\ell=1}^{m} E_\ell)^{1/d} \,\right)\,\right)  \,\notag\\
	&\stackrel{n \rightarrow \infty}{\longrightarrow} 0 \label{eq:converge_1}\;,
\end{align}
where the last inequality follows from Lemma \ref{lem:order_stat}. By the assumption that $f$ has open support and $\|\nabla f\|$ and $\|H_f\|$ is bounded almost everywhere, this convergence holds for almost every $x$.

For the second term in \eqref{eq:entropy3},
let $\tilde{T}^{(m)}_{\alpha} = \tS^{(\infty)}_\alpha - \tS^{(m)}_\alpha$ and
we claim that $\tilde{T}^{(m)}_{\alpha}$ converges to 0 in distribution by the following lemma.
\begin{lemma}
    \label{lem:tail}
    Assume $m \to \infty$ as $n \to \infty$ and $k\geq 3$ , %and $\tilde{T}^{(m)}_{\alpha,i}$ defined as in~\eqref{eq:T},
    then % we know that:
    \begin{eqnarray}
    \lim_{n \to \infty} \E \|\, \tilde{T}^{(m)}_{\alpha} \,\| = 0
    \end{eqnarray}
    for any $\alpha \in \{0,1,2\}$. Hence $(\tilde{T}^{(m)}_{0},\tilde{T}^{(m)}_{1},\tilde{T}^{(m)}_{2})$ converges to $(0,0,0)$ in distribution.
\end{lemma}
%Since $\tilde{S}^{(m)}_{\alpha,i} = \tilde{S}^{(\infty)}_{\alpha,i} - \tilde{T}^{(m)}_{\alpha,i}$ for any $\alpha \in \{0,1,2\}$, we know that
This implies that  $(\tilde{S}^{(m)}_{0},\tilde{S}^{(m)}_{1},\tilde{S}^{(m)}_{2})$ converges to $(\tilde{S}^{(\infty)}_{0},\tilde{S}^{(\infty)}_{1},\tilde{S}^{(\infty)}_{2})$ in distribution, i.e.,
\begin{eqnarray}
%\|\,(\tilde{S}^{(m)}_{0,i},\tilde{S}^{(m)}_{1,i},\tilde{S}^{(m)}_{2,i}) - (\tilde{S}^{(\infty)}_{0,i},\tilde{S}^{(\infty)}_{1,i},\tilde{S}^{(\infty)}_{2,i})\,\|_{TV}
d_{\rm TV}(\tS^{(m)},\tS^{(\infty)}) \stackrel{n \rightarrow \infty}{\longrightarrow} 0 \label{eq:converge_2}\;,
\end{eqnarray}

Combine~\eqref{eq:converge_1} and ~\eqref{eq:converge_2} in \eqref{eq:entropy3}, this implies the desired claim.

%Notice that $(\xi_k \sum_{l=1}^k E_\ell, \tilde{S}^{(\infty)}_{0,i}, \tilde{S}^{(\infty)}_{1,i}, \tilde{S}^{(\infty)}_{2,i})$ is a random variable depends only on $k$ (not on $n$ or underlying pdf $f(x)$). So we know that $(2nf(x) \rho_{k,i}, S_{0,i}, S_{1,i}, S_{2,i})$ converges to $(\xi_k \sum_{l=1}^k E_\ell, \tilde{S}^{(\infty)}_{0,i}, \tilde{S}^{(\infty)}_{1,i}, \tilde{S}^{(\infty)}_{2,i})$ in distribution.

We next prove the upper bound on the variance, following the technique from \cite[Section 7.3]{BD16}.
 For the usage of Efron-Stein inequality, we need a second set of i.i.d. samples $\{X'_1, X'_2, \dots, X'_n\}$. For simplicity, denote $\hH = \hH_{kLNN}^{(n)}(X)$ be the kLNN estimate base on original sample $\{X_1, \dots, X_n\}$ and $\hH^{(i)}$ be the kLNN estimate based on $\{X_1, \dots, X_{i-1}, X'_i, X_{i+1}, \dots X_n\}$. Then Efron-Stein theorem states that
\begin{eqnarray}
\Var \left[ \hH \right] \leq 2 \sum_{j=1}^n  \mathbb{E} \left[\, \left( \hH - \hH^{(j)} \right)^2 \,\right] \;.\label{eq:efron_stein}
\end{eqnarray}

Recall that
\begin{eqnarray*}
	\hH = \frac1n \sum_{i=1}^n \Big\{ \underbrace{h\big(\,(c_d nf(X_i))^{1/d} Z_{k,i}, S_{0,i}, S_{1,i}, S_{2,i})\,\big) - \log f(X_i)}_{\equiv H_i}  \Big\} \;,
\end{eqnarray*}
where  $h : \mathbb{R}^d \times \mathbb{R} \times \mathbb{R}^d \times \mathbb{R}^{d \times d} \to \mathbb{R}$ is defined as
\begin{align}
&h(t_1, t_2, t_3, t_4) = \nonumber\\
&\;\;d \log \|t_1\| + d \log (2 \pi) - \log c_d - \log t_2  + \frac{1}{2} \log \left(\, \det\left(\, \frac{t_4}{t_2} - \frac{t_3 t_3^T}{t^2_2}\,\right) \,\right)+ \frac{1}{2} t_3^T (t_4 - t_3 t_3^T)^{-1} t_3 \;. 
\end{align}
 Similarly, we can write $\hH^{(j)} = \frac1n \sum_{i=1}^n H_i^{(j)}$ for any $j \in \{1, \dots, n\}$. Therefore, the difference of $\hH$ and $\hH^{(j)}$ can be bounded by:
\begin{eqnarray}
 \hH - \hH^{(j)} = \frac{1}{n} \sum_{i=1}^n \left(\, H_i - H_i^{(j)} \,\right) \;.
\end{eqnarray}

Notice that $H_i$ only depends on $X_i$ and its $m$ nearest neighbors, so $H_i - H_i^{(j)} = 0$ if none of $X_j$ and $X'_j$ are in $m$ nearest neighbor of $X_i$. If we denote $Z_{i,j} = \mathbb{I} \{X_j \textrm{ is in } m \textrm{ nearest neighbor of } X_i\}$, then $H_i = H_i^{(j)}$ if $Z_{i,j}+Z_{i,j'} = 0$. According to \cite[Lemma 20.6]{BD16}, since $X$ has a density, with probability one, $\sum_{i=1}^n Z_{i,j} \leq m \gamma_d$, where $\gamma_d$ is the minimal number of cones of angle $\pi/6$ that can cover $\mathbb{R}^d$, which only depends on $d$. Similarly, $\sum_{i=1}^n Z_{i,j'} \leq m \gamma_d$. If we denote $S = \{i: Z_{i,j} + Z_{i,j'} > 0\}$, the cardinality of $S$ satisfy $|S| \leq 2 m \gamma_d$. Therefore, we have $\hH - \hH^{(j)} = \frac{1}{n} \sum_{i \in S} \left(\, H_i - H_i^{(j)} \,\right)$. By Cauchy-Schwarz inequality, we have
\begin{eqnarray}
\mathbb{E} \left[\, \left( \hH - \hH^{(j)} \right)^2 \,\right] &=& \mathbb{E} \left[\, \frac{1}{n^2} \left(\, \sum_{i \in S} \left(\, H_i - H_i^{(j)} \,\right) \,\right)^2\,\right] \,\notag\\
&\leq& \mathbb{E} \left[\, \frac{|S|}{n^2} \sum_{i \in S} \left(\, H_i - H_i^{(j)} \,\right)^2\,\right] \,\notag\\
&=& \frac{|S|}{n^2} \sum_{i \in S} \mathbb{E} \left[\, \left(\, H_i - H_i^{(j)}\,\right)^2\,\right] \,\notag\\
&\leq& \frac{2|S|}{n^2} \sum_{i \in S} \left(\, \mathbb{E} \left[\, H_i^2\,\right] + \mathbb{E} \left[\, (H_i^{(j)})^2\,\right] \,\right) \;.\label{eq:h-h_j}
\end{eqnarray}

Notice that $H_i$'s and $H_i^{(j)}$'s are identically distributed, so we are left to compute $\mathbb{E} \left[\, H_1^2 \,\right]$. Conditioning on $X_1 = x$, similarly to~\eqref{eq:entropy2}, we have
\begin{eqnarray}
\E \left[\, (H_1 + \log f(x))^2 | X_1 = x\,\right] &=& \E \left[\, {h}^2( (c_d nf(x))^{1/d} Z_{k,i}, S_{0,1}, S_{1,i}, S_{2,i}) \,\right] \,\notag\\
& \longrightarrow & B^{(2)}_{k,d}\;,
\end{eqnarray}
as $n \to \infty$, where $B^{(2)}_{k,d} \equiv \E\left[\, {h}^2 \left(\,\xi_k \left(\, \sum_{\ell=1}^k E_\ell \,\right)^{1/d}, \tilde{S}^{(\infty)}_{0}, \tilde{S}^{(\infty)}_{1}, \tilde{S}^{(\infty)}_{2} \,\right) \,\right]$. Therefore,
\begin{eqnarray}
\E \left[\, H_1^2 | X_1 = x\,\right] &=& B^{(2)}_{k,d} - 2 \log f(x) \E \left[\, H_1 | X_1 = x\,\right] - (\log f(x))^2 \,\notag\\
&=& B^{(2)}_{k,d} - 2 \log f(x) B_{k,d} + (\log f(x))^2 \;.
\end{eqnarray}
Take expectation over $X_1$, we obtain:
\begin{eqnarray}
\E [H_1^2] &=& \E_{X_1} \left[\, \lim_{n \to \infty} \E \left[\, H_1^2 | X_1 \,\right] \,\right] = \E_{X_1} \left[\, B^{(2)}_{k,d} - 2 \log f(X_1) B_{k,d} + (\log f(X_1))^2 \,\right] \,\notag\\
&=& B^{(2)}_{k,d} + 2 H(X) B_{k,d} + \int f(x) (\log f(x))^2 dx < +\infty\;,
\end{eqnarray}
where the last inequality comes from the assumption that $\int f(x) (\log f(x))^2 dx < +\infty$. Combining with~\eqref{eq:efron_stein} and~\eqref{eq:h-h_j}, we have
\begin{eqnarray}
\Var \left[ \hH \right] \leq 2 \sum_{j=1}^n  \mathbb{E} \left[\, \left( \hH - \hH^{(j)} \right)^2 \,\right] \leq\frac{4|S|}{n} \sum_{i \in S} \left(\, \mathbb{E} \left[\, H_i^2\,\right] + \mathbb{E} \left[\, (H_i^{(j)})^2\,\right] \,\right) \leq \frac{8|S|^2 C_2}{n} \leq \frac{32m^2 \gamma_d^2 C_2}{n} \;,
\end{eqnarray}
where $C_2$ is the upper bound for $\E [H_1^2]$. Take $m = O(\log n)$ then the proof is complete.

% ------------------------------------------------------------------------------------------------------------------------------------------------

\subsection{Proof of Lemma~\ref{lem:tail}}

Firstly, since $|\xi_i| = 1$, we can upper bound the expectation of $\E \|\, \tilde{T}^{(m)}_{\alpha,i} \,\|$ by:
\begin{eqnarray}
\E \|\, \tilde{T}^{(m)}_{\alpha,i} \,\| &=& \E \, \Big\|\, \sum_{j=m+1}^{\infty} \frac{\xi_j^{(\alpha)} (\sum_{\ell=1}^j E_{\ell})^{\alpha/d}}{(\sum_{\ell=1}^k E_\ell)^{\alpha/d} } \exp\{- \frac{(\,\sum_{\ell=1}^j E_\ell\,)^{2/d}}{2(\,\sum_{\ell=1}^k E_\ell\,)^{2/d}}\} \Big\|\, \,\notag\\
&\leq& \sum_{j=m+1}^{\infty} \E \,\Big\|\, \frac{ \xi_j^{(\alpha)} (\sum_{\ell=1}^j E_{\ell})^{\alpha/d} }{(\sum_{\ell=1}^k E_\ell)^{\alpha/d} } \exp\{- \frac{(\,\sum_{\ell=1}^j E_{\ell}\,)^{2/d}}{2(\,\sum_{\ell=1}^k E_\ell\,)^{2/d}}\} \,\Big\| \,\notag\\
&=& \sum_{j=m+1}^{\infty} \E \,\Big|\, \frac{(\sum_{\ell=1}^j E_{\ell})^{\alpha/d}}{(\,\sum_{\ell=1}^k E_\ell\,)^{\alpha/d}} \exp\{- \frac{(\,\sum_{\ell=1}^j E_{\ell}\,)^{2/d}}{2(\,\sum_{\ell=1}^k E_\ell\,)^{2/d}}\} \,\Big|\;. \label{eq:eq1}
\end{eqnarray}
Notice that the expression is a function of $(\sum_{\ell=1}^j E_{\ell}/\sum_{\ell=1}^k E_{\ell})^{1/d} \equiv R_j$ for $j > m$, we will identify the distribution of $R_j$ first. For any fixed $j \geq k$, let $T_k = \sum_{\ell=1}^k E_{\ell}$ and $T_{j-k} = \sum_{\ell=k+1}^j E_{\ell}$, such that $R_j = ((T_k + T_{j-k})/T_k)^{1/d}$. Notice that $T_k$ is the summation of $k$ i.i.d. standard exponential random variables, so $T_k \sim \textit{ Erlang } (k,1)$. Similarly, $T_{j-k} \sim \textit{ Erlang }(j-k,1)$. Also $T_k$ and $T_{j-k}$ are independent. Recall that the pdf of $\textit{ Erlang }(k, \lambda)$ is given by $f_{k,\lambda}(x) = \lambda^k x^{k-1} e^{-\lambda x}/(k-1)!$ for $x \geq 0$. Therefore, the CDF of $R_j$ is given by:
\begin{eqnarray}
F_{R_j}(t) &=& \Pr \{R_j \leq t\} = \Pr\{ (\frac{T_k + T_{j-k}}{T_k})^{1/d} \leq t\} = \Pr \{\frac{T_{j-k}}{T_k} \leq t^d - 1\} \,\notag\\
&=& \int_{x \geq 0} \frac{x^{k-1} e^{-x}}{(k-1)!} \left(\, \int_{y=0}^{(t^d-1)x} \frac{y^{j-k-1}e^{-y}}{(j-k-1)!} dy \,\right) dx \,\notag\\
&=& \int_{x \geq 0} \frac{x^{k-1} e^{-x}}{(k-1)!} \left(\, 1 - \sum_{\ell=0}^{j-k-1} \frac{1}{\ell!} x^{\ell} (t^d-1)^{\ell} e^{-x(t^d-1)} \,\right) dx \,\notag\\
&=& 1 - \sum_{\ell=0}^{j-k-1} \left(\, \int_{x \geq 0} \frac{x^{k-1}e^{-x}}{(k-1)!} \frac{1}{\ell!} x^{\ell}(t^d-1)^{\ell} e^{-x(t^d-1)} dx\,\right) \,\notag\\
&=& 1 - \sum_{\ell=0}^{j-k-1} \left(\, \frac{(t^d-1)^{\ell}}{(k-1)! \ell!} \int_{x \geq 0} x^{k-1+\ell} e^{-xt^d} dx\,\right) \,\notag\\
&=& 1 - \sum_{\ell=0}^{j-k-1} \frac{(t^d-1)^{\ell}}{(k-1)! \ell!} \,(k-1+\ell)! \,t^{-d(k-1+\ell)} \,\notag\\
&=& 1- \sum_{\ell=0}^{j-k-1} {k-1+\ell \choose \ell} \, t^{-d(k-1)} (1-t^{-d})^{\ell} \;,
\end{eqnarray}
for $t \in [1,+\infty)$. Given the CDF of $R_j$, each term in~\eqref{eq:eq1} is upper bounded by:
\begin{eqnarray}
&& \E \,\Big|\, \frac{(\sum_{\ell=1}^j E_{\ell})^{\alpha/d}}{(\,\sum_{\ell=1}^k E_\ell\,)^{\alpha/d}} \exp\{- \frac{(\,\sum_{\ell=1}^j E_{\ell}\,)^{2/d}}{2(\,\sum_{\ell=1}^k E_\ell\,)^{2/d}}\} \,\Big| = \E_{R_j} \Big|\, t^{\alpha} e^{-t^2} \,\Big| \leq \E_{R_j} \left[\, t^{2} e^{-t^2} \,\right] \,\notag\\
&=& \int_{t=1}^{\infty} t^2 e^{-t^2} dF_{R_j}(t)= t^2 e^{-t^2} F_{R_j}(t) \Big|_{1}^{\infty} - \int_{t=1}^{\infty} F_{R_j}(t) d(t^2 e^{-t^2}) \,\notag\\
&=& - \int_{t=1}^{\infty} (2te^{-t^2} - 2t^3 e^{-t^2}) F_{R_j}(t) dt = \int_{t=1}^{\infty} 2t(t^2-1)e^{-t^2} F_{R_j}(t) dt \;. \label{eq:eq2}
\end{eqnarray}
Therefore, in order to establish an upper bound for~\eqref{eq:eq1}, we need an upper bound for $F_{R_j}(t)$. Here we will consider two cases depending on $t$. If $t > (j/2k)^{1/d}$, we just use the trivial upper bound $F_{R_j}(t) < 1$. If $1 \leq t \leq (j/2k)^{1/d}$, since $t^d \geq 1$, we have:
\begin{eqnarray}
F_{R_j}(t) = 1- \sum_{\ell=0}^{j-k-1} {k-1+\ell \choose \ell} \, t^{-d(k-1)} (1-t^{-d})^{\ell} \leq 1- \sum_{\ell=0}^{j-k-1} {k-1+\ell \choose \ell} \, t^{-dk} (1-t^{-d})^{\ell} \;.
\end{eqnarray}
Notice that ${k-1+\ell \choose \ell} \, t^{-dk} (1-t^{-d})^{\ell}$ is the pmf of negative binomial distribution ${\rm NB}(k, 1-t^{-d})$. Therefore, $F_{R_j}(t) \leq \Pr\{ X \geq j-k\}$, where $X \sim {\rm NB}(k, 1-t^{-d})$. The mean and variance of $X$ are given by $\E[X] = (1-t^{-d})k/(1-(1-t^{-d})) = (t^d-1)k$ and ${\rm Var}(X) = (1-t^{-d})k/(1-(1-t^{-d}))^2 = (t^{2d}-t^d)k$. Therefore, by Chebyshev inequality, the tail probability is upper bounded by:
\begin{eqnarray}
\Pr\{ X \geq j-k \} \leq \frac{{\rm Var}(X)}{(j-k-\E[X])^2} = \frac{(t^{2d}-t^d)k}{(j-k-(t^d-1)k)^2} = \frac{(t^{2d}-t^d)k}{(j-t^dk)^2} \leq 4t^{2d}k/j^2 \;,
\end{eqnarray}
here we use the fact that $t \leq (j/2k)^{1/d}$ so $j - t^dk > j/2$. Therefore, $F_{R_j}(t) \leq 4t^{2d}k/j^2$ for $t > (j/2k)^{1/d}$. Combine the two cases and plug into~\eqref{eq:eq2}, we obtain:
\begin{eqnarray}
&& \E \,\Big|\, \frac{(\sum_{\ell=1}^j E_{\ell})^{\alpha/d}}{(\,\sum_{\ell=1}^k E_\ell\,)^{\alpha/d}} \exp\{- \frac{(\,\sum_{\ell=1}^j E_{\ell}\,)^{2/d}}{2(\,\sum_{\ell=1}^k E_\ell\,)^{2/d}}\} \,\Big| = \int_{t=1}^{\infty} 2t(t^2-1)e^{-t^2} F_{R_j}(t) dt \,\notag\\
&\leq& \int_{t=1}^{(j/2k)^{1/d}} 2t(t^2-1) e^{-t^2} \frac{4t^{2d}k}{j^2} dt + \int_{(j/2k)^{1/d}}^{\infty} 2t(t^2-1) e^{-t^2} dt \,\notag\\
&\leq& \frac{8k}{j^2} \int_{t=1}^{\infty} t^{2d+3} e^{-t^2} dt + 2 \int_{(j/2k)^{1/d}}^{\infty} t^3 e^{-t^2} dt \,\notag\\
&\leq& \frac{8k C_{d}}{j^2} + 2 \left(\, -\frac{1}{2}e^{-t^2}(t^2+1) \Big|_{(j/2k)^{1/d}}^{\infty}\,\right) \,\notag\\
&=& \frac{8k C_{d}}{j^2} + e^{-(j/2k)^{2/d}} ((\frac{j}{2k})^{2/d}+1) \;,
\end{eqnarray}
where $C_d = \int_{t=1}^{\infty} t^{2d+3} e^{-t^2} dt$ is a constant only depend on $d$. Therefore, we can see that
\begin{eqnarray}
\E \,\Big|\, \frac{(\sum_{\ell=1}^j E_{\ell})^{\alpha/d}}{(\,\sum_{\ell=1}^k E_\ell\,)^{\alpha/d}} \exp\{- \frac{(\,\sum_{\ell=1}^j E_{\ell}\,)^{2/d}}{2(\,\sum_{\ell=1}^k E_\ell\,)^{2/d}}\} \,\Big| = O(1/j^2).
\end{eqnarray}
 So
\begin{eqnarray}
\E \|\, \tilde{T}^{(m)}_{\alpha,i} \,\| &\leq& \sum_{j=m+1}^{\infty} \E \,\Big|\, \frac{(\sum_{\ell=1}^j E_{\ell})^{\alpha/d}}{(\,\sum_{\ell=1}^k E_\ell\,)^{\alpha/d}} \exp\{- \frac{(\,\sum_{\ell=1}^j E_{\ell}\,)^{2/d}}{2(\,\sum_{\ell=1}^k E_\ell\,)^{2/d}}\} \,\Big| \to 0\;. \label{eq:eq1}
\end{eqnarray}
given $m \to \infty$ as $n \to \infty$.

% ------------------------------------------------------------------------------------------------------------------------------------------------------
\section{Proof of Theorem \ref{thm:unify}}
\label{sec:proof_unify}

The proposed estimator is a solution to a maximization problem $\ha = \arg\max_a \cL_{X_i}(f_{a,X_i})$.
From \cite{Loa96}  we know that the maximizer is a fixed point of a series of non-linear equations of the form
\begin{eqnarray*}
	&&\sum_{j\neq i}  \frac{(X_j-X_i)^{ \otimes \alpha }}{\rho_{k,i}^{\alpha}} K\Big( \frac{X_j-X_i}{\rho_{k,i}} \Big) \\
	&&=\;\;n \,\rho_{k,i}^d \,e^{a_0} \int  \frac{(u-X_i)^{ \otimes \alpha }}{\rho_{k,i}^{\alpha}} K\Big( \frac{u-X_i}{\rho_{k,i}} \Big)
	e^{\<u-x,a_1\> + \cdots + a_p[(u-x),\cdots,(u-x)]}   \frac{1}{\rho_{k,i}^{d}} \,du
\end{eqnarray*}
for all $\alpha\in[p]$
where the superscript $\otimes\alpha$ indicates the $\alpha$-th order tensor product.
From the proof of Theorem \ref{thm:unbiased}, specifically \eqref{eq:converge_1} and \eqref{eq:converge_2},
we know that the left-hand side converges to a value that only depends on $k,d$ and $K$.
Let's denote it by $S_\alpha(k)\in\reals^{d^\alpha}$.
We make a change of variables $\ta_0=a_0+d\log \rho_{k,i} + \log n$
and $\ta_\alpha=a_\alpha/\rho_{k,i}^\alpha$ for $\alpha\neq 0$.
Then, in the limit of growing $n$, the above equations can be rewritten as
\begin{eqnarray}
	S_\alpha(k,d,K) &=& e^{\ta_0} F_\alpha(d,K,\ta_1,\ldots,\ta_p) \;,
\end{eqnarray}
for some function $F_\alpha$. Notice that the dependence on the underlying distribution vanishes in the limit,
and the fixed point $\ta$ only depends on $k$, $p$, $d$, and $K$. The desired claim follows from the fact that
the estimate is $\lim_{n\to \infty} \hf_n(X_i)=\lim_{n\to \infty}e^{\ha_0} =  \lim_{n\to \infty}A_{k,d,p,K}/(n \rho_{k,i}^d)=
f(X_i)A_{k,d,p,K}C_d \lim_{n}1 / (C_d n \rho_{k,i}^d f(X_i)) = f(X_i)A_{k,d,p,K}C_d /\sum_{\ell=1}^k E_\ell $,  and plugging in the entropy estimator
$\hH(X) \to E_{X_i}[-\log f(X_i)]+B_{k,d,p,K} $.

In the case of the KL estimator, it happens that
$S_0=k$ and $F_0(d)=C_d$ such that $e^{\ta_0}=k/C_d$,
$e^{\ha_0} = f(X_i) k/(C_d \rho_{k,i}^d f(X_i)n)$ and $B_{k,d,p,K} = -\log k + E[\log (\sum_{\ell=1}^k E_\ell)] = -\log k + \phi(k)$.

\section*{Acknowledgement}
This work is supported by NSF SaTC award CNS-1527754, NSF CISE award CCF-1553452,
NSF CISE award CCF-1617745.
We thank the anonymous reviewers for their constructive feedback.

% ---------------------------------------------------------------------------------------------------------------------------------
%\newpage
\bibliographystyle{plain}
{\small
\bibliography{local}}

\end{document}